\documentclass[useAMS,usenatbib]{mn2e}
\usepackage{longtable}
\usepackage{graphicx}
\usepackage{txfonts}
\usepackage{natbib}

\title[Data Mining for Dwarf Novae]{Data Mining for Dwarf Novae in SDSS, GALEX and Astrometric Catalogues}
\author[P. Wils et al.]{Patrick Wils$^{1}$, Boris T. G\"ansicke$^{2}$, Andrew J. Drake$^{3}$, John Southworth$^{2}$\\
$^{1}$Vereniging voor Sterrenkunde, Belgium, email: patrickwils@yahoo.com \\
$^{2}$Department of Physics, University of Warwick, Coventry CV4 7AL, UK \\
$^{3}$California Institute of Technology, 1200 E. California Blvd, CA 91225, USA}
\begin{document}

\date{Accepted 2009 ???? ??. Received 2009 ???? ??}

\pagerange{\pageref{firstpage}--\pageref{lastpage}} \pubyear{2009}

\maketitle

\label{firstpage}

\begin{abstract}
By cross matching blue objects from SDSS with GALEX and the
astrometric catalogues USNO-B1.0, GSC2.3 and CMC14, 64 new dwarf nova
candidates with one or more observed outbursts have been
identified. 14 of these systems are confirmed as cataclysmic
variables through existing and follow-up spectroscopy.
A study of the amplitude
distribution and an estimate of the outburst frequency of these new
dwarf novae and those discovered by the Catalina Real-time Transient
Survey (CRTS) indicates that besides systems that are faint because
they are farther away, there also exists a population of intrinsically
faint dwarf novae with rare outbursts.
\end{abstract}

\begin{keywords}
stars: novae, cataclysmic variables --
stars: dwarf novae --
\end{keywords}

%
%________________________________________________________________

\section{Introduction}

One of the astrophysically interesting evolutionary phases a close
binary star can go through, is that of a cataclysmic variable (CV), in
which a main-sequence star filling its Roche lobe loses mass to a
white dwarf.  In the case the white dwarf has no or only a weak
magnetic field, an accretion disc forms around it.  Because of thermal
instabilities in the accretion disc some of these systems, the dwarf
novae, occasionally outburst and become several magnitudes brighter
for a few days to a few weeks at most
\citep{meyer+meyer-hofmeister84-1, osaki96-1, lasota01-1}.  
  Theoretical studies of the orbital period distribution of these
  dwarf novae predicted a large number of systems with an orbital
  period near the period minimum \citep{kolb93-1, howelletal01-1}.
  Until recently the predicted accumulation of CVs near the period minimum
  was not what was observed \citep{kolb+baraffe99-1, willemsetal05-1}.
  Furthermore there is a discrepancy between the theoretical period
  minimum of $\approx$ 70 minutes and the observed one between 80 and
  86 minutes \citep{kolb+baraffe99-1}.  The Sloan Digital Sky Survey
(SDSS) has led to the spectroscopic identification of a large sample
of faint CVs \citep[see][and earlier papers in the
  series]{szkodyetal09-1}. Follow-up studies reveal that this faint
SDSS CV population differs in many aspects from the previously known
brighter systems, and that its period distribution is in closer
agreement with the theoretical predictions \citep{gaensickeetal09-2},
 although the discrepancy between the observed and the
  theoretically predicted orbital period minimum remains.  Also the
  observed ratio of short to long period CVs still disagrees with the
  model predictions \citep{aungwerojwit06-1,pretoriusetal07-1,pretorius+knigge08-1}.  
While SDSS has made already a substantial
contribution to observational population studies of CVs, the
completeness of the currently known CV sample remains rather
uncertain.

Prior to SDSS, a large fraction of all CVs in general, and the
majority of dwarf novae in particular, have been discovered
serendipitously by their large amplitude variability, favouring those
systems with frequent outbursts \citep{gaensicke05-1}. The majority of
dwarf novae are expected to have evolved to short periods near the
period minimum, and to have a low mass transfer rate. Consequently, they
have infrequent outbursts, and the probability for a serendipitous
discovery is lower than for systems with longer orbital periods and
higher mass transfer rates. Historically, a large fraction of these
serendipitous discoveries were made by amateur observers, which
implies that the current sample of dwarf novae is biased towards
relatively bright objects. In addition, it is very difficult to assess
the spatial and temporal coverage of their searches for transient
objects. The Catalina Real-time Transient
Survey\footnote{http://voeventnet.caltech.edu/feeds/Catalina.shtml}
\citep[CRTS;][] {drakeetal09-1}, based on data from the Catalina Sky Survey (CSS),
is now routinely discovering dwarf
novae.  Because this is a much more systematic survey, which also goes to fainter magnitudes, 
it is expected that it will eventually provide strong
constraints on the total number and the magnitude distribution of
dwarf novae.

Here, we cross-match a number of large surveys to find faint
outbursting dwarf novae, and make use of CRTS light curves to compare
the properties of the previously known dwarf novae, those identified
spectroscopically by SDSS, and the ones discovered in this paper.

%__________________________________________________________________

\section{Search for Dwarf Novae}
\label{search}
\begin{figure}
\centering
\includegraphics[width=\columnwidth]{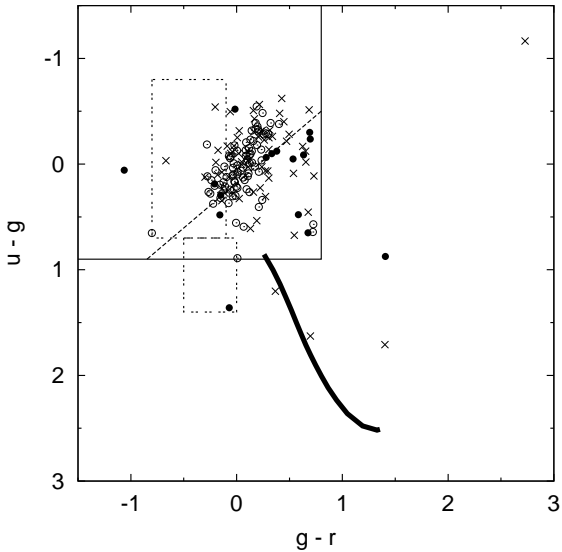}
\caption{SDSS colours for a sample of 163 known dwarf novae.
  Open circles represent systems with an orbital period below the period gap,
  filled circles those with a period above the period gap.
  Crosses indicate the systems for which the orbital period is not known.
  The full lines denote the cutoffs used for selecting candidate dwarf novae.
  For matching with astrometric catalogues, candidates needed to lie
  above the dashed line as well.  The curved line represents the
  stellar locus as determined by \citet{richardsetal02-1}.  For
  illustration purposes only, the exclusion boxes for white dwarfs
  (top) and A stars (bottom) from \citet{richardsetal02-1} are drawn with
  dotted lines.  The outlier point in the top right corner is for
  CSS090102:132536+210037.  It is possible that the object entered an
  eclipse during the time SDSS cycled through the different
  passbands. }
\label{Cutoff}
\end{figure}

To constrain the colours of candidate dwarf novae, SDSS photometry was
obtained for a list of known dwarf novae taken from the AAVSO Variable
Star Index (VSX)\footnote{http://www.aavso.org/vsx}. A sample of 163
objects was selected, presented in Table~\ref{DNSample}.  Excluding
outliers, this led to the following conditions:
\begin{equation}
\label{colourcond}
u - g < 0.9, \\
g - r < 0.8, \\
r - i < 0.8, \\
i - z < 1.0. \\
\end{equation}
Lower limits on the colours were not deemed necessary.
The cutoffs for the $u-g$ and $g-r$ colours are illustrated in Fig.~\ref{Cutoff}.

\begin{table*}
\begin{center}
\caption{Known dwarf novae with SDSS data.  An asterisk after the $r$
  magnitude indicates the object was in outburst at the time of the
  SDSS photometry.  The dwarf nova subtypes are defined as follows:
  UG: U Gem without subclassification; UGSS: SS Cyg subtype; UGZ: Z
  Cam subtype; UGSU: SU UMa subtype, UGWZ: WZ Sge subtype; E:
  eclipsing; ZZ: ZZ Ceti type white dwarf pulsations.  The
    orbital period, taken from the online catalogue of \citet{ritter+kolb03-1}, is given in hours.  The
    recovery method refers to the four sub-steps from
    Fig.\,\ref{Flowchart}: A (match with GALEX), B (SDSS variability), C (no USNO
    counterpart) and D (matching with
    astrometric catalogues).  The full table is available in the
  online edition.}
\label{DNSample}
\begin{tabular}{lclllrrrrc}
\hline
Name                  & SDSS  &     Type & $P_{orb}$ & \multicolumn{1}{c}{$r$} & $u-g$ & $g-r$ & $r-i$ & $i-z$ & Method \\
\hline
TY Psc                    & J012539.35+322308.6 & UGSU   & 1.64 & 13.05* & 0.37 & -0.08 & -0.28 & -0.34 & D \\
SDSS J013132.39-090122.3  & J013132.39-090122.2 & UG+ZZ  & 1.36 & 18.42  & 0.02 & -0.12 & -0.18 & 0.13 &  \\
SDSS J013701.06-091235.0  & J013701.06-091234.8 & UGSU   & 1.33 & 18.45  & 0.34 & 0.24 & 0.38 & 0.28 &  \\
SDSS J015151.87+140047.2  & J015151.87+140047.2 & UG     & 1.98 & 20.00  & -0.29 & 0.30 & 0.41 & 0.45 &  \\
SDSS J031051.66-075500.3  & J031051.66-075500.3 & UGSU   & 1.60 & 15.74* & 0.26 & -0.27 & -0.16 & -0.22 & D \\
CSS081107:033104+172540   & J033104.44+172540.2 & UG     &      & 19.33  & 0.09 & 0.54 & 0.45 & 0.41 & D \\
SDSS J033449.86-071047.8  & J033449.86-071047.8 & UGSU   & 1.90 & 14.85* & 0.28 & -0.25 & -0.18 & -0.19 & D \\
SDSS J033710.91-065059.4  & J033710.91-065059.4 & UG     &      & 19.72  & 0.10 & -0.18 & -0.25 & -0.17 &  \\
CSS080303:073921+222454   & J073921.17+222453.5 & UG     &      & 22.40  & 0.06 & 0.27 & 0.27 & -0.37 &  \\
\multicolumn{1}{c}{...} & \multicolumn{1}{c}{...} & \multicolumn{1}{c}{...} & \multicolumn{5}{c}{...} \\
\hline
\end{tabular}
\end{center}
\end{table*}

\begin{figure*}
\centering
\includegraphics[angle=-90,width=\textwidth]{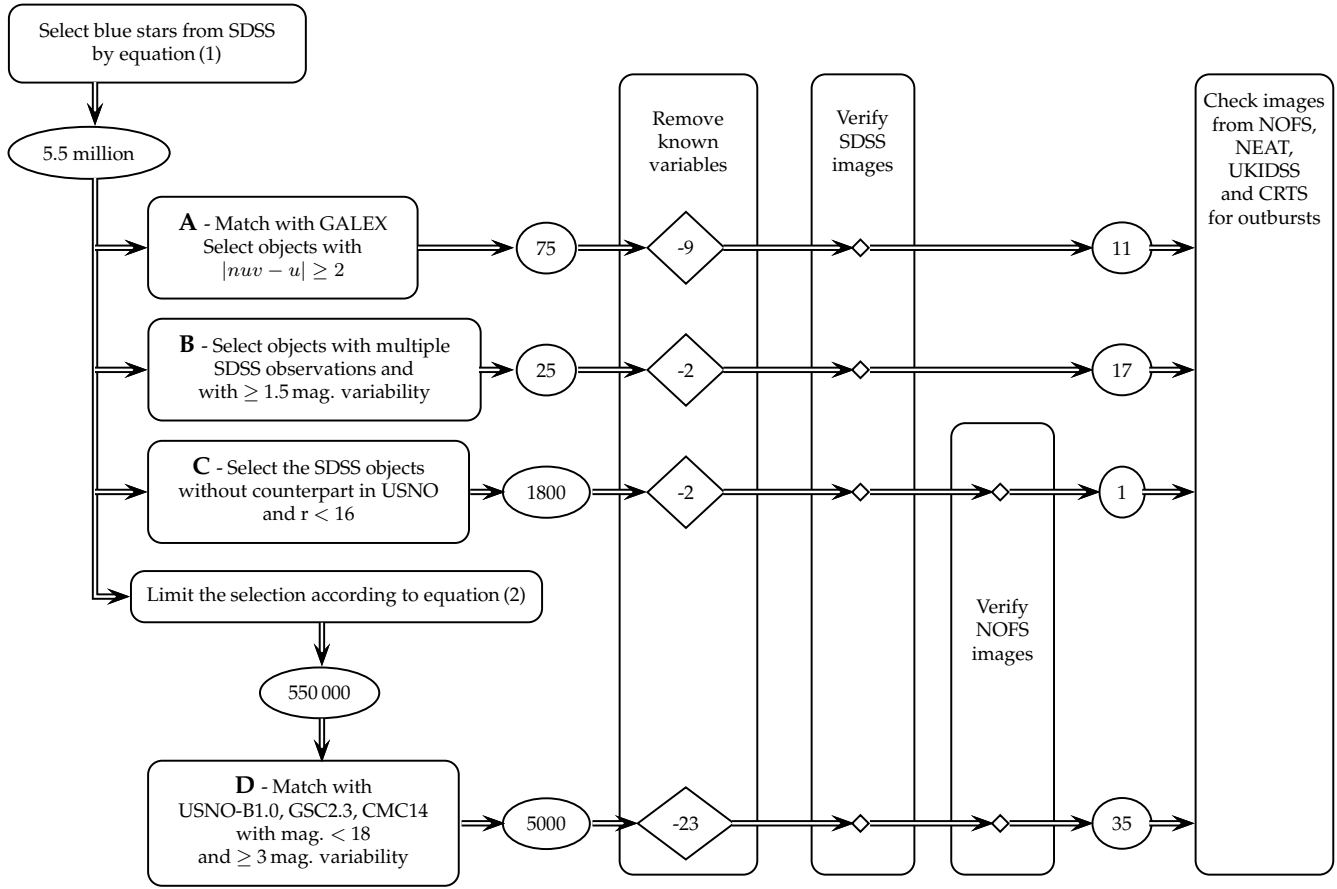}
\caption{  Schematic presentation of the search procedure.
    Numbers in ellipses indicate the number of objects selected in the
    previous step.  Bold capital letters A-D indicate the different
    methods explained in detail in the text. }
\label{Flowchart}
\end{figure*}

A summary of the search procedure is given as a flow chart in Fig.~\ref{Flowchart}. 
Our initial master list of objects are all point sources within SDSS
Data Release~7 \citep{abazajian09-1} satisfying
equation~(\ref{colourcond}). These objects were then matched to repeat
observations within SDSS and to GALEX objects \citep{martinetal05-1},
adopting a search radius of 1~arc second.  For this the predefined
cross-matching tables in SDSS and GALEX CasJobs \citep{li+thakar08-1}
and \citet{budavarietal09-1} were used.

The master list was also matched to the entries in the astrometric
catalogues USNO-B1.0 \citep{monetetal03-1}, GSC2.3
\citep{laskeretal08-1} and CMC14 \citep{cmc14}, using VizieR
\citep{ochsenbeinetal00-1}.  In this case a matching radius of 2 arc
seconds was used for USNO-B1.0 and GSC2.3, and 1 arc second for CMC14,
but in most cases the resulting matches were much closer than this
radius.  To further limit the number of candidates to match against
these astrometric catalogues, the following more stringent condition
on the colours was applied:
\begin{equation}
\label{extracond}
  (u-g) + 0.85 (g-r) < 0.18.
\end{equation}
This condition was found by optimising the number of known dwarf
  novae kept in the sample, while 90\% of the blue stars were
  excluded.  Condition (\ref{extracond}) still retained more than 60\%
  of the original sample of 163 objects, and thus maintained the best
  chances to find new dwarf novae, while the amount of work involved
  in checking candidates was reduced tenfold.  A bias toward short
  period dwarf novae is however introduced in this way: of the systems
  with an orbital period below the period gap 73\% (58 out of 79)
  satisfy condition (\ref{extracond}), while only 31\% (5 out of 16)
  of the systems above the period gap do.  This can be explained by
  the fact that short period systems are mostly white dwarf dominated,
  while in longer period systems, the red dwarf will be larger, and
  therefore the combined colour will be redder.  In the $(u - g, g -
  r)$ diagram, most CVs seem to cluster around a line with a slope
  similar to the one in condition (\ref{extracond}) and almost
  perpendicular to the main sequence \citep[see Fig.~\ref{Cutoff} and
    also Fig.~6 from ][]{gaensickeetal09-2}.  

\begin{table*}
\begin{center}
\caption{Candidate and confirmed dwarf novae. The SDSS magnitudes are
  from DR7 \citep{abazajian09-1}, ultraviolet magnitudes have been
  obtained from GALEX \citep{martinetal05-1}.  An asterisk after the
  $nuv$ and the $u$ magnitude indicates that the GALEX and SDSS
  measurements were taken during outburst, respectively.  The
  approximate observed amplitude of variability is provided in the $r$
  band when available, otherwise the brightest and faintest
    observed magnitude are given, irrespective of passband.  The
  discovery method refers to the four sub-steps from
  Fig.\,\ref{Flowchart}: A (match with GALEX), B (SDSS variability), C (no USNO
  counterpart) and D (matching with astrometric
  catalogues).  Identification spectra are available for 16 objects,
  magnitudes at the time of the spectroscopy are provided where
  available.  }
\label{CVs}
\begin{tabular}{ccllccccccl}
\hline
SDSS                & $fuv$ & \multicolumn{1}{c}{$nuv$}  & \multicolumn{1}{c}{$u$}    & $g$   & $r$   & $i$   & $z$   & Mag. range & Method & Spectrum \\
\hline
J013645.81-193949.1 & 20.36 & 20.52  & 20.40  & 20.27 & 20.21 & 20.09 & 19.68 & 16.1-20.2 & D &  \\
J015237.83-172019.3 &       &        & 21.43  & 21.38 & 21.43 & 21.30 & 20.99 & 16.0-21.4 & D &  \\
J032015.29+441059.3 &       &        & 18.43  & 18.77 & 18.43 & 18.20 & 18.02 & 14.7-18.4 & B &  \\
J064911.48+102322.1 &       &        & 17.61* & 17.26 & 17.20 & 17.24 & 17.29 & 17.2-18.9 & B &  \\
J073208.11+413008.7 & 17.07 & 17.09* & 20.45  & 20.77 & 20.44 & 20.23 & 19.84 & 15.6-20.4 & B & SDSS ($g=20.7$) \\
J073758.55+205544.5 & 20.24 & 20.01  & 17.19* & 17.39 & 17.23 & 17.28 & 17.27 & 17.2-20.0 & B &  \\
J074500.58+332859.6 & 20.73 & 20.48  & 22.47  & 22.21 & 22.25 & 21.71 & 21.21 & 18.0-22.2 & B &  \\
J074859.55+312512.6 & 16.46 & 16.39* & 15.89* & 15.74 & 15.94 & 16.01 & 16.11 & 14.9-17.8 & B &  \\
J075107.50+300628.4 & 19.02 & 19.62  & 19.57  & 19.76 & 19.81 & 19.83 & 19.74 & 15.1-19.8 & D &  \\
J075117.00+100016.2 & 18.70 & 18.35  & 17.93  & 18.52 & 18.39 & 18.41 & 18.33 & 14.5-18.4 & D &  \\
J075713.81+222253.0 & 17.88 & 17.93* & 20.98  & 21.26 & 21.18 & 21.06 & 21.40 & 17.0-21.2 & A &  \\
J080033.86+192416.5 &       &        & 19.55  & 19.74 & 19.72 & 19.71 & 19.58 & 15.9-19.7 & B &  \\
J080306.99+284855.8 &       & 21.12  & 16.48* & 16.28 & 16.42 & 16.58 & 16.70 & 15.8-20.4 & B &  \\
J081030.45+091111.7 &       &        & 19.74* & 19.98 & 19.84 & 19.72 & 17.62 & 19.8-22.8 & B &  \\
J081408.42+090759.1 &       &        & 20.57  & 20.69 & 20.46 & 19.96 & 19.42 & 14.4-20.5 & D &  \\
J081529.89+171152.5 &       &        & 22.08  & 21.84 & 21.99 & 21.87 & 21.73 & 17.2-22.0 & D &  \\
J082648.28-000037.7 &       &        & 20.63  & 21.02 & 20.85 & 20.82 & 20.59 & 16.0-20.9 & D &  \\
J083132.41+031420.7 & 21.20 & 21.08  & 17.37* & 17.15 & 17.44 & 17.66 & 17.83 & 17.1-21.2 & D & GMOS ($r=17.6$) \\
J083508.99+600643.9 &       &        & 21.86  & 21.66 & 21.41 & 21.25 & 21.55 & 18.3-21.4 & B &  \\
J084011.95+244709.8 &       & 22.14  & 19.07* & 19.03 & 19.37 & 19.48 & 18.94 & 19.4-21.2 & B &  \\
J084108.10+102536.2 & 20.90 & 21.12  & 19.85  & 20.24 & 20.14 & 20.00 & 20.08 & 15.8-20.1 & D &  \\
J091147.02+315101.8 & 20.80 & 20.69  & 15.35* & 15.17 & 15.51 & 15.76 & 15.96 & 15.4-20.0 & D &  \\
J091242.18+620940.1 &       & 20.04  & 18.73  & 18.81 & 18.73 & 18.44 & 17.86 & 14.5-18.7 & D & SDSS ($g=19.5$) \\
J091741.29+073647.4 &       & 22.20  & 21.77  & 21.51 & 21.39 & 21.51 & 21.15 & 18.2-21.4 & D &  \\
J092620.42+034542.3 & 18.40 & 17.87* & 19.67  & 19.91 & 19.80 & 19.80 & 19.86 & 17.0-19.8 & A &  \\
J092809.84+071130.5 & 22.07 & 21.40  & 15.58* & 15.28 & 15.47 & 15.67 & 15.84 & 15.5-21.0 & D & SDSS ($g=14.7$) \\
J093946.03+065209.4 &       & 22.21  & 18.00* & 17.74 & 18.00 & 18.17 & 18.38 & 17.4-22.0 & B & SDSS ($g=17.7$) \\
J100243.11-024635.9 &       &        & 21.18  & 21.38 & 21.52 & 21.24 & 20.47 & 16.3-21.5 & D &  \\
J100516.61+694136.5 &       &        & 19.80  & 19.40 & 18.92 & 18.60 & 18.26 & 17.9-21.0 & D & SDSS ($g=18.4$) \\
J105333.76+285033.6 & 18.27 & 18.03* & 20.23  & 20.16 & 19.79 & 19.60 & 19.44 & 17.2-20.6 & A &  \\
J112003.40+663632.4 & 22.13 & 21.53  & 15.81* & 15.62 & 15.90 & 16.12 & 16.32 & 15.9-21.0 & D & SDSS ($g=21.1-18.2$) \\
J120054.13+285925.2 &       & 21.49  & 16.96* & 16.71 & 17.00 & 17.20 & 17.35 & 17.0-21.0 & D &  \\
J124328.27-055431.0 & 19.51 & 19.88* & 23.21  & 22.74 & 22.52 & 22.29 & 21.81 & 19.5-22.5 & A &  \\
J124602.02-202302.4 & 18.60 & 19.31  & 18.37  & 18.57 & 18.43 & 18.38 & 18.29 & 15.5-18.4 & D & 6dFGS \\
J124719.03+013842.6 & 18.56 & 18.21* & 20.61  & 20.69 & 20.92 & 21.05 & 21.13 & 17.6-20.9 & A &  \\
J131432.10+444138.7 & 19.89 & 19.70  & 15.10* & 15.08 & 15.27 & 15.43 & 15.61 & 15.3-20.0 & D &  \\
J132040.96-030016.7 & 19.89 & 19.44  & 26.10  & 25.71 & 21.69 & 21.69 & 21.13 & 19.4-21.7 & A &  \\
J132715.28+425932.8 & 17.76 & 17.90* & 20.94  & 20.79 & 20.52 & 20.34 & 20.19 & 17.9-20.5 & A &  \\
J133820.56+041807.4 &       &        & 18.53* & 19.18 & 17.99 & 18.60 & 16.74 & 18.0-21.4 & B &  \\
J141029.09+330706.2 &       &        & 23.25  & 22.44 & 22.24 & 22.24 & 22.56 & 19.8-22.2 & B &  \\
J142414.20+105759.8 &       &        & 18.41* & 16.45 & 15.95 & 16.86 & 16.41 & 15.9-19.0 & B &  \\
J142953.56+073231.2 & 21.65 & 21.19  & 21.00  & 21.21 & 20.80 & 20.32 & 19.81 & 16.9-20.8 & D &  \\
J151109.79+574100.3 & 17.97 & 17.74* & 20.89  & 21.24 & 21.18 & 21.04 & 20.88 & 17.3-21.2 & A &  \\
J152124.38+112551.9 &       &        & 20.02* & 19.40 & 19.28 & 19.28 & 19.16 & 19.3-21.7 & B & GMOS ($g=21.8$) \\
J153457.24+505616.8 &       & 22.09  & 21.94  & 22.01 & 21.78 & 21.72 & 21.04 & 17.0-21.8 & D &  \\
J154357.66+203942.1 & 22.36 & 21.25  & 16.93* & 16.58 & 16.74 & 16.88 & 17.01 & 15.6-21.0 & D &  \\
J154652.70+375415.2 & 16.39 & 16.46  & 16.01* & 16.03 & 16.27 & 16.46 & 16.66 & 16.3-21.0 & D &  \\
J154817.56+153221.2 & 17.51 & 17.33* & 21.55  & 21.75 & 21.65 & 21.69 & 21.32 & 16.5-21.7 & A &  \\
J155030.38-001417.3 & 18.93 & 19.01  & 21.91  & 22.01 & 21.60 & 21.04 & 20.77 & 17.1-21.6 & A &  \\
J155540.19+364643.1 & 20.66 & 20.68  & 20.48  & 20.80 & 20.90 & 20.72 & 20.52 & 16.0-20.9 & D & ACAM \\
J161027.61+090738.4 & 20.95 & 20.65  & 20.19  & 20.10 & 20.06 & 20.26 & 20.03 & 15.0-20.1 & D & SDSS ($g=20.2$) \\
J161442.43+080407.9 &       & 23.41  & 21.44  & 21.24 & 21.41 & 20.95 & 21.35 & 16.8-21.4 & D &  \\
J162520.29+120308.7 & 20.99 & 20.09  & 18.45  & 18.48 & 18.41 & 18.24 & 17.69 & 14.5-18.4 & D & SDSS ($g=19.9$) \\
J162558.18+364200.6 & 21.73 & 22.04  & 21.49  & 21.68 & 21.60 & 21.54 & 21.89 & 17.1-21.6 & D &  \\
J162900.55+341022.0 & 21.61 & 21.89  & 21.16  & 21.22 & 21.20 & 21.12 & 21.09 & 16.0-21.2 & D & ACAM \\
J164705.07+193335.0 &       &        & 21.91  & 21.95 & 21.79 & 21.68 & 21.34 & 16.7-21.8 & D &  \\
\hline
\multicolumn{11}{l}{...continued on next page.}
\end{tabular}
\end{center}
\end{table*}
\begin{table*}
\begin{center}
\contcaption{}
\begin{tabular}{ccllccccccl}
\hline
SDSS                & $fuv$ & \multicolumn{1}{c}{$nuv$}  & \multicolumn{1}{c}{$u$}    & $g$   & $r$   & $i$   & $z$   & Mag. range & Method & Spectrum \\
\hline
J170145.85+332339.5 &       &        & 21.23  & 21.58 & 21.40 & 21.31 & 21.65 & 18.8-21.4 & B & GMOS ($g=22.2$) \\
J170810.31+445450.7 & 20.40 & 20.89  & 20.78  & 20.83 & 20.78 & 20.84 & 20.87 & 14.0-20.8 & D & ACAM \\
J171202.95+275411.0 &       &        & 21.24  & 21.39 & 20.94 & 20.51 & 20.58 & 17.3-20.9 & D & GMOS ($g=21.0$) \\
J174839.77+502420.3 &       &        & 23.24  & 22.30 & 23.02 & 23.78 & 24.15 & 15.3-23.0 & C &  \\
J191616.53+385810.6 &       &        & 17.94* & 17.67 & 17.56 & 17.49 & 17.44 & 15.5-19.3 & D &  \\
J205931.86-070516.6 & 21.33 & 20.64  & 20.85  & 20.75 & 21.03 & 21.24 & 21.56 & 17.4-21.0 & D &  \\
J212025.17+194156.3 & 16.12 & 16.17* & 21.85  & 21.81 & 21.81 & 21.65 & 22.72 & 16.1-21.8 & A &  \\
J223854.51+053606.8 &       & 21.58  & 21.52  & 21.52 & 21.51 & 21.52 & 20.82 & 17.4-21.5 & D &  \\
\hline
\end{tabular}
\end{center}
\end{table*}

Objects were then selected as candidate transients when $|nuv - u|\geq
2$ for the GALEX cross-matches (method ``A''), or when all of the $ugriz$ magnitudes differed
  by at least 1.5 mag.\ in the case of multiple SDSS imaging runs
  (method ``B''), or when it had no
counterpart in the USNO cross-matching table from SDSS CasJobs
  (method ``C'').  Further candidates were selected from objects that
had an entry brighter than mag. 18 in one of the astrometric
catalogues, with at least a 3 mag.\ difference with respect to the
SDSS $r$ value (method ``D'').  The reason for the latter
constraint is that because USNO-B1.0 and GSC2.3 are constructed from
single images at a given epoch, possible outbursts cannot be confirmed
on a second image from around that time.  Given the limiting magnitude
of about 20 for these catalogues, outburst detections fainter than
mag. 18 might be questionable.  Since dwarf novae in outburst are
nearly white ($B-V \approx 0$), no colour-correction was made for the
filters used to obtain the catalogue magnitudes.

\begin{table*}
\begin{center}
\caption{Observed outbursts of the new dwarf novae.
 Sources for the data are:
\citet[CMC14;][]{cmc14}, Digitized Sky Survey (DSS),
\citet[CRTS;][]{drakeetal09-1},
\citet[GALEX;][]{martinetal05-1},
Near Earth Asteroid Tracking survey (NEAT),
United States Naval Observatory, Flagstaff Station (NOFS),
\citet[PM2000;][]{ducourantetal06-1}
\citet[SDSS;][]{abazajian09-1}
and \citet[UKIDSS;][]{lawrenceetal07-1}.
The table is available in full in the online edition.}
\label{Outbursts}
\begin{tabular}{cl}
\hline
SDSS                & Outbursts \\
\hline
J013645.81-193949.1 & Oct 1997 (NOFS), Dec 2000 (NEAT), Nov 2004 (CMC14) \\
J015237.83-172019.3 & Sep 1978 (NOFS) \\
J032015.29+441059.3 & Nov 1994 (NOFS), Dec 2005 (SDSS) \\
J064911.48+102322.1 & Nov 2006 (SDSS) \\
J073208.11+413008.7 & Mar 1953 (NOFS), Dec 2001 (NEAT), Nov 2003 (SDSS), Feb 2007 (GALEX + CRTS) + 3 other outbursts (CRTS) \\
J073758.55+205544.5 & Jan 2002 (SDSS), Dec 2002 (NEAT), Feb 2006 (CRTS) \\
J074500.58+332859.6 & See text \\
J074859.55+312512.6 & Dec 2001 (SDSS + NEAT), + 4 bright and 6 fainter outbursts (CRTS) \\
J075107.50+300628.4 & Mar 2003 (CMC14 + NEAT), Mar 2005 (CRTS), Jan 2006 (CRTS), Jan 2007 (CRTS), Dec 2007 (CRTS) \\
... & ... \\
\hline
\end{tabular}
\end{center}
\end{table*}

In this way some 6500 transient candidates remained in total 
(taking into account that some candidates were selected by different methods).  
  Among them 36 known dwarf novae from the original sample in
  Table\,\ref{DNSample} were recovered.  The repartition among the
  different methods is given in Fig.~\ref{Flowchart}.  The recovered
  sample viewed as a whole contained 22\% of the original sample of
  systems below the period gap, and 19\% of the systems above the
  period gap.  For method ``D'' only, these fractions are respectively
  14\% and 6\%.  So, although this method is biased towards systems
  below the period gap, this bias is reflected less in the total number
  of recovered objects.  The 36 known dwarf novae consisted of 15
confirmed SU~UMa stars, one WZ~Sge star (AL~Com) and also 7 dwarf
novae recently discovered by CRTS.  An outburst to $r=13.7$ (CMC14,
April 2003) of the suspected large amplitude eruptive variable EL~UMa
was also found, confirming it to be a dwarf nova.  The first outburst
of this system was detected by \citet{pesch+sanduleak87-1} in May
1981.  In addition to those 36 known dwarf novae, also the nova
  CT~Ser was recovered.  The USNO-B1.0 B1 and R1 images date from June
  1950, about two and a half years after its 1948 outburst which
  reached at least magnitude 8.5 \citep{lohmann49-1}.  At the time of
  the images CT~Ser was still brighter (around mag. 13) than in
  quiescence (mag. 16).

SDSS images were then inspected to make sure that the remaining
candidates were indeed stars, since a large fraction of the selected
objects turned out to be parts of galaxies, or foreground stars to
galaxies.  In the latter case the photometry in the astrometric
catalogues cannot be trusted. Also some stars in the glare of a nearby
bright star were excluded.  The candidates were then examined on
images of the United States Naval Observatory, Flagstaff Station
(NOFS)\footnote{http://www.nofs.navy.mil/data/fchpix/} to exclude
image artefacts in the compilation of the USNO-B1.0 and GSC2.3
catalogues.  Images of the Near Earth Asteroid
Tracking
(NEAT)\footnote{http://skyview.gsfc.nasa.gov/skymorph/skymorph.html}
survey were also examined to confirm the variability of some of the dubious
candidates, and to look for additional outbursts of the real
transients.  Two possible outbursts were also found in data of the
UKIRT Infrared Deep Sky Survey \citep[UKIDSS;][]{lawrenceetal07-1,
  warrenetal07-1}.  As a large number of images were inspected
visually, we do not claim that this search is exhaustive.  Finally for
the remaining real transients, photometric data from CRTS were
collected (dating back to late 2004), resulting in the detection of
additional outbursts\footnote{The CRTS light curves compiled for this
  analysis are made available at
  http://nesssi.cacr.caltech.edu/catalina/20010604/}.  One of the
reasons the CRTS system did not trigger a transient alert when some of these
objects were observed in outburst was that the CRTS pipeline compares
its photometric data with SDSS and USNO-B1.0. When a dwarf nova has
been observed in outburst in one of these surveys it cannot reach the
threshold required for a transient detection.

As a result, a total of 64 new candidate dwarf novae were found, which
are listed in Table~\ref{CVs} together with their SDSS magnitudes.
Table~\ref{Outbursts} gives a list of the observed outbursts, where an
outburst is defined as an observation to within one magnitude of the
brightest detection from Table~\ref{CVs}.

There is always the possibility that some of the outbursting dwarf
novae candidates found in this study, are in reality optically violent
variable quasars (OVV) quasars.  Similar to dwarf novae, they can show
outbursts up to 6 magnitudes \citep{abraham+carrara98-1}, but these
outbursts generally last more than three months, much longer than a
typical dwarf nova outburst.  However, because of the limited data
available for some of the outbursts from this paper, a distinction is
not always possible.  Unlike faint dwarf novae, OVV quasars and
blazars in general are often also detected in radio wavelengths.  None
of the candidates dwarf novae presented in Table~\ref{CVs} are however
known to have a radio counterpart.  In addition, only three known
quasars were found (SDSS\,J095315.74+441123.6, 2QZ\,J095551.3-020428
and SDSS\,J103705.22+463128.4, all by matching with astrometric
catalogues), as opposed to 36 known dwarf novae.

\begin{figure}
\centering
\includegraphics[width=\columnwidth]{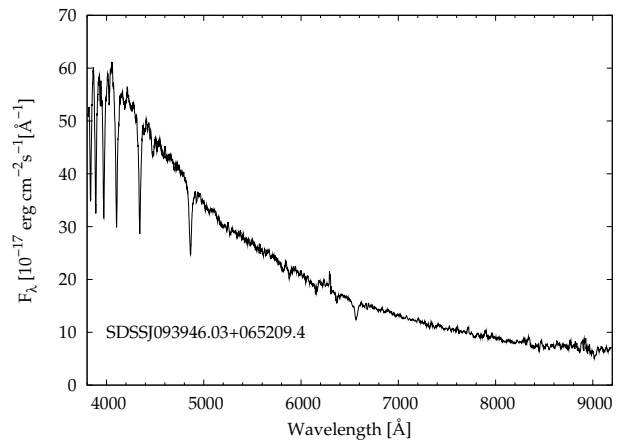}
\caption{ SDSS DR3 spectrum \citep{abazajian05-1} of
    SDSS\,J093946.03+065209.4 in outburst.}
\label{f-sdss}
\end{figure}

\begin{figure}
\centering
\includegraphics[width=\columnwidth]{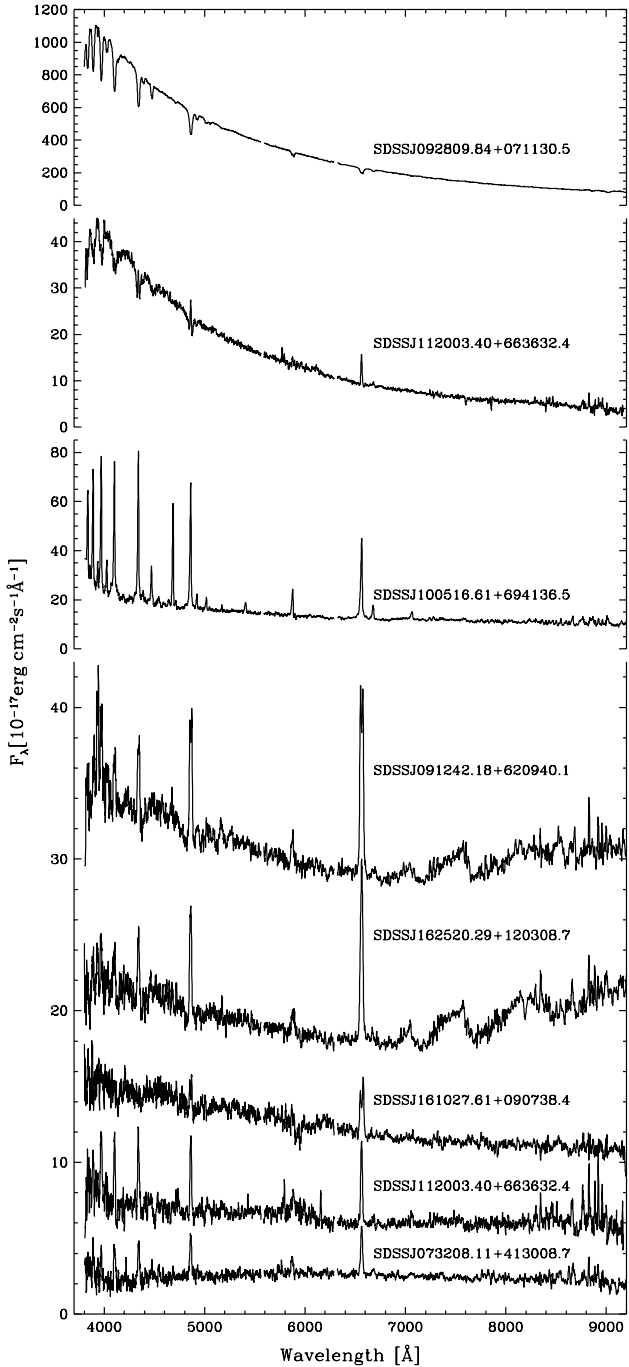}
\caption{ SDSS DR7 \citep{abazajian09-1} spectra of seven CV
    candidates from Table\,\ref{CVs}. The spectra in the bottom
    panel are offset (top to bottom) by 25, 15, 10, 5 and 0 flux
    units. SDSS\,J112003.40+663632.4 was observed in quiescence
    (bottom panel) and on the decline from an outburst (2nd panel from
    top).}
\label{f-dr7}
\end{figure}

\begin{figure}
\centering
\includegraphics[width=\columnwidth]{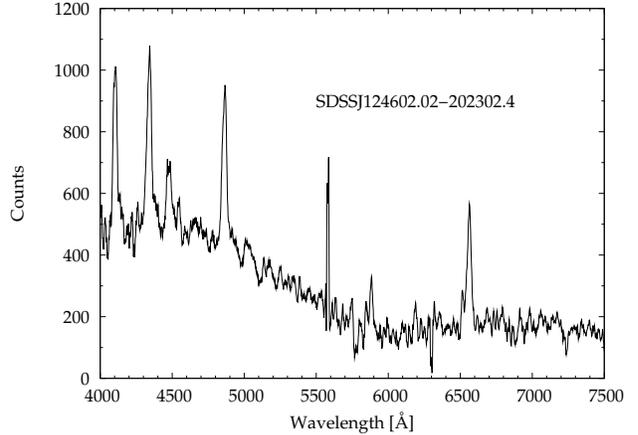}
\caption{ Spectrum of SDSS\,J124602.02-202302.4 from the 6dF
    Galaxy Survey \citep{jonesetal05-1,jonesetal09-1}.}
\label{f-6df}
\end{figure}

%
%______________________________________________________________

\section{Identification spectroscopy}
\label{s-spectroscopy}

Identification spectra are available for 16 of the 64 dwarf nova
candidates: one from SDSS DR3 (Fig.\,\ref{f-sdss},
\citealt{abazajian05-1}), 7 within SDSS DR7 (Fig.\,\ref{f-dr7},
\citealt{abazajian09-1}), and one from the 6dF Galaxy
Survey\footnote{http://www-wfau.roe.ac.uk/6dFGS} (Fig.\,\ref{f-6df},
\citealt{jonesetal05-1, jonesetal09-1}).  We obtained spectra for
the other 7 objects using GMOS on Gemini-North and ACAM on the
William Herschel Telescope (Figs.\,\ref{f-acam} and \ref{f-gemini}).
Details of the GMOS and ACAM observations are briefly outlined in
Sect.\,\ref{s-spectroscopy} before discussing details of individual
systems in Sect.\,\ref{s-individual_systems}.

\subsection{WHT/ACAM}

Spectroscopic observations were carried out in service mode in 2009
June, using the new
ACAM\footnote{http://www.ing.iac.es/Astronomy/instruments/acam/index.html}
spectrograph on the 4.2\,m William Herschel Telescope (WHT). The
default 400 lines\,mm$^{-1}$ grism was used, giving a wavelength
coverage of 4000--9000\,\AA. The detector was an EEV 4096$\times$2048
pixel CCD, binned by a factor of two in both dimensions to give a
spatial resolution of 0.5$^{\prime\prime}$\,px$^{-1}$ and a reciprocal
dispersion of approximately 7\,\AA\,px$^{-1}$ throughout the
wavelength range.

The data were reduced using optimal extraction \citep{horne86-1} as
implemented in the {\sc pamela}\footnote{{\sc pamela} and {\sc molly}
  were written by TRM and can be found at
  http://www.warwick.ac.uk/go/trmarsh} code
\citep{marsh89-1}. Wavelength calibration was performed using
copper-argon and copper-neon arc lamp spectra taken immediately before
or after the science observations, but is very approximate bluewards
of 5850\,\AA\ due to the lack of measurable emission lines in this
region. From fitting Gaussian functions to emission lines from the
night sky and from the arc lamps, we estimate the resolution of these
spectra to be approximately 18\,\AA.

\begin{figure}
\centering
\includegraphics[angle=-90,width=\columnwidth]{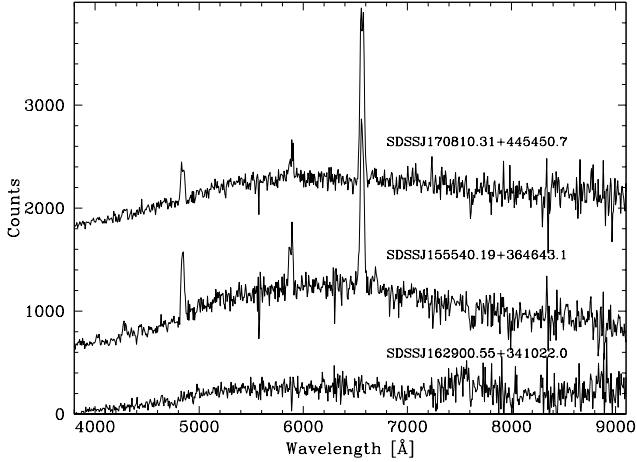}
\caption{WHT/ACAM spectra for three of the new dwarf novae presented
  in this paper.}
\label{f-acam}
\end{figure}

The reduced spectra (Fig.\,\ref{f-acam}) were corrected for telluric
absorption using spectra of the flux standard star
BD\,+26$^\circ$2606. Due to problems arising from the poor throughput
at wavelengths below $\simeq4500$\,\AA, we decided not to apply any flux
calibration.

\subsection{Gemini/GMOS}

\begin{figure}
\centering
\includegraphics[width=\columnwidth]{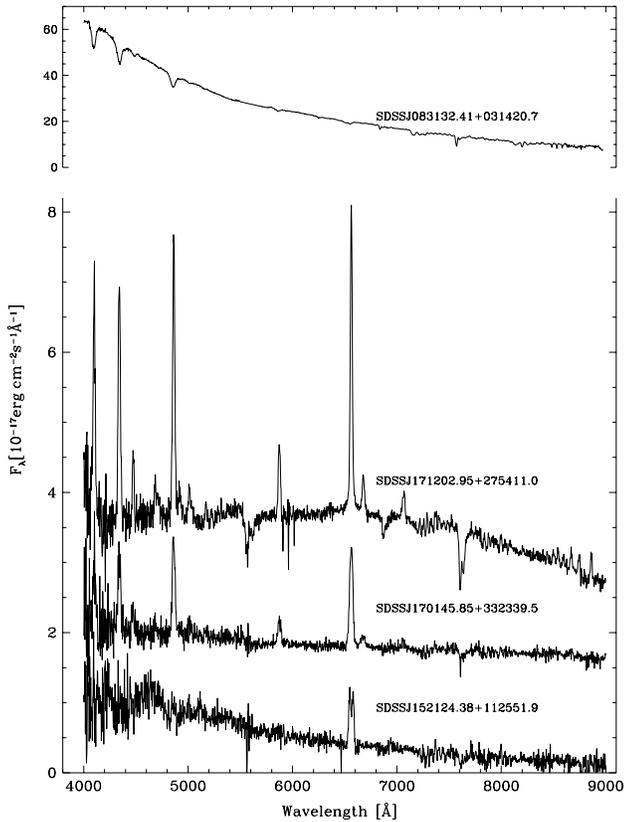}
\caption{Gemini/GMOS spectra for four of the new dwarf novae presented
  in this paper. SDSS\,J083132.41+031420.7 was observed during
  outburst.}
\label{f-gemini}
\end{figure}

Spectroscopic observations were performed with the Gemini-South
telescope between 2008 June and 2009 April, with the GMOS-S instrument
and the R150 grating. We used a central wavelength setting of 740\,nm
and a slit of width 1$^{\prime\prime}$. The detector of this
spectrograph is a mosaic of three 2048$\times$4068 pixel CCDs abutted
on their long edges, and long-slit spectra are dispersed over all
three CCDs. We reduced the spectra from CCD2 and CCD3 individually, in
the same way as for the ACAM data, and ignoring CCD1 which covers the
spectral region redwards of 9300\,\AA. The spectra were then
flux-calibrated using observations of LTT\,6248 and the results from
the different CCDs combined to form a summed spectrum for each object
covering 3500-9200\,\AA. Severe fringing is apparent redwards of
7000\,\AA. The wavelength calibration of CCD3 (3500--5550\,\AA) is
only approximate due to the very small number of usable arc lines.
The final spectra have a reciprocal dispersion of 3.5\,\AA\ and a
resolution of roughly 15\,\AA. The reduced spectra are shown in
Fig.\,\ref{f-gemini}.

\section{Details on Individual Systems}
\label{s-individual_systems}

All the new systems with spectra, and some other systems with
  noteworthy aspects are briefly discussed in this section.
Additional cross-identifications for objects from Table~\ref{CVs}
which are listed in other catalogues, but were not yet identified as
dwarf nova candidates, are given in Table~\ref{Crossids}.

\begin{table*}
\begin{center}
\caption{Cross identifications for the new dwarf nova candidates.} \vskip2mm
\label{Crossids}
\begin{tabular}{lll}
\hline
SDSS                & Other ID                & Reference \\
\hline
J092620.42+034542.3 & GUVV J092620.4+034541.8 & \citet{welshetal05-1} \\
J124328.27-055431.0 & GUVV J124328.2-055431.7 & \citet{welshetal05-1} \\
J124602.02-202302.4 & 6dFGS g1246019-202302   & \citet{jonesetal05-1,jonesetal09-1} \\
J131432.10+444138.7 & 2RXP J131432.9+444144 \\
J132040.96-030016.7 & 1RXS J132041.2-030010   & \citet{vogesetal99-1} \\
J132715.28+425932.8 & GUVV J132715.2+425932.1 & \citet{welshetal05-1} \\
J162900.55+341022.0 & NSV 20657               & \citet{hovanissian78-1,hovanissian82-1} \\
\hline
\end{tabular}
\end{center}
\end{table*}

\subsection{Systems with spectra}

Among the 16 systems for which spectroscopy has been obtained 13
  have spectra consistent with being dwarf novae, one appears to be a
  polar or intermediate polar (SDSS\,J100516.61+694136.5), and two
  have spectra that are unusual for CVs (SDSS\,J093946.03+065209.4 and
  SDSS\,J162900.55+341022.0). SDSS\,J093946.03+065209.4 is the only
  pre-DR7 CV candidate that has not been listed by
  \citet{szkodyetal09-1} and the previous papers of that series, for
  the reasons outlined below. The CVs with DR7 spectroscopy will be
  included in \citet{szkodyetal09-2}.

\newcounter{itemno}
\begin{list}{~~~(\roman{itemno})}{\usecounter{itemno}}

\item \textit{SDSS\,J073208.11+413008.7.} The SDSS spectrum exhibits a
  rather red slope (Fig.\,\ref{f-dr7}), and one of the three SDSS
  photometric data sets shows a very unusual $ugriz$ spectral energy
  distribution. The CRTS light curve of this object shows a high duty
  cycle, implying frequent outbursts. An alternative possibility is
  that the system is eclipsing.

\item \textit{SDSS\,J083132.41+031420.7.} We observed this object
  with Gemini when it was in a bright state ($r\simeq17.6$).  The
  spectrum is typical of a dwarf nova in outburst, with broad Balmer
  absorption lines from an optically thick accretion disc
  (Fig.\,\ref{f-gemini}).

\item \textit{SDSS\,J091242.18+620940.1.} Both the white dwarf and the
  M-dwarf companion are visible in the SDSS spectrum
  (Fig.\,\ref{f-dr7}), implying that this system has a low accretion
  rate.

\item \textit{SDSS\,J092809.84+071130.5.} The SDSS spectrum was
  obtained during a bright state ($g=14.7$), and is typical of a dwarf
  nova in outburst, similar to SDSS\,J083132.41+031420.7
  (Fig.\,\ref{f-dr7}).

\item \textit{SDSS\,J093946.03+065209.4.} This object was classified
  as a subdwarf by \citet{eisensteinetal06-1}.  Its spectrum is
  actually more reminiscent of a star of spectral type A
  (Fig.\,\ref{f-sdss}), and is very unusual for a CV.  This is
    probably also the reason why it was not included in the list of
    CVs from SDSS~DR3 \citep{szkodyetal04-1}.  However both NEAT and
  SDSS have it bright (around magnitude 18) the first half of March
  2002 on two nights each, while it is faint (mag. 22) on two nights
  in 2006 in SDSS, and invisible on NEAT images on 30 other nights,
  including images taken at the end of February 2002.  CRTS observed
  other outbursts to mag. 17.4 (May 2005), 18.2 (February 2006), 19.1
  (January 2008) and 17.8 (June 2008), but did not detect it most of
  the time either.  The SDSS spectroscopic fluxes (from a year after
  the photometry) agree almost spot-on with the photometry, suggesting
  another outburst in March 2003.  This may be a CV with a very low
  inclination, where the Balmer lines from the accretion disc are only
  marginally Doppler-broadened.

\item \textit{SDSS\,J100516.61+694136.5.} With a strong HeII 4686 line
  (Fig.\,\ref{f-dr7}), it is likely a magnetic CV, i.e. a polar or an
  intermediate polar.  It is listed in the 2MASS catalogue with
  $J=16.8$.

\item \textit{SDSS\,J112003.40+663632.4.} The SDSS imaging found
  SDSS\,J1120+6636 at $r=15.9$ on MJD51661. SDSS spectroscopy of
  SDSS\,J1120+6636 was obtained on 5 different nights. On three
  occasions, the system was observed in quiescence at $r\simeq21$ (MJD
  54464, 54483, 54485), and its spectrum exhibits narrow, moderately
  strong Balmer emission lines superimposed on a weak continuum
  (Fig.\,\ref{f-dr7}). Spectra obtained on MJD 54495 and 54498 showed
  the system on the decline from an outburst at $r\simeq18.2$ and
  $\simeq19.4$. The average SDSS spectrum available in DR7 is
  dominated by the data obtained on MJD 54495, and show a steep blue
  continuum with rather broad Balmer absorption lines, superimposed by
  weak emission lines with only the central core visible, as
  expected for an accretion disc in/close to outburst. CRTS detected a
  single additional outburst on MJD 53746.

\item \textit{SDSS\,J124602.02-202302.4.} The 6dFGS spectrum
(Fig.\,\ref{f-6df}) is reminiscent of a quiescent dwarf nova, with
broad Balmer emission lines.

\item \textit{SDSS\,J152124.38+112551.9.} This is an example of a CV
  with a white-dwarf dominated optical spectrum and a steep Balmer
  line decrement (Fig.\,\ref{f-gemini}), resembling closely the class
  of intrinsically faint CVs with periods within the 80--86\,min
  period minimum spike that have been identified from SDSS
  spectroscopy \citep{gaensickeetal09-2}.  The spectrum implies a low
  temperature in the accretion flow, which again most likely means a
  low accretion rate.

\item \textit{SDSS\,J155540.19+364643.1.}  The ACAM spectrum
(Fig.\,\ref{f-acam}) is typical of a quiescent dwarf nova, with strong
Balmer and He\,I emission lines.

\item \textit{SDSS\,J161027.61+090738.4.} Based on the white-dwarf
  dominated SDSS spectrum (Fig.\,\ref{f-dr7}), we suggest that this
  object is a WZ\,Sge type dwarf nova, with one outburst detected in
  June 1998 on NEAT images. While preparing the manuscript,
  SDSS\,1610+0907 has entered a superoutburst, and follow-up
  photometry suggests a superhump period of
  $\simeq82$\,min\footnote{http://ooruri.kusastro.kyoto-u.ac.jp/pipermail/vsnet-alert/2009-August/002989.html}.

\item \textit{SDSS\,J162520.29+120308.7.} The SDSS spectrum is
  dominated by emission from the white dwarf and the M-dwarf companion
  star, superimposed by broad emission lines (Fig.\,\ref{f-dr7}). The
  spectrum is similar to that of RZ~Leo.  The 2MASS catalogue lists
  the object at $J=16.6$ and $K_s=15.5$. The CRTS light curve is very
  unusual, exhibiting besides one outburst a modulation of the
  quiescent magnitude with a $\sim1$\,mag amplitude on time scales of
  $\sim500$ days. This is suggestive of stellar activity cycles on the
  companion star.

\item \textit{SDSS\,J162900.55+341022.0.} This is the only dwarf nova
  candidate where follow-up spectroscopy fails to detect Balmer
  emission lines (Fig.\,\ref{f-acam}). The SDSS co-ordinates are
  consistent with those of NSV\,20657, a bright blue object ($B=17.9$,
  $U-B=-0.6$ \citep{hovanissian78-1,hovanissian82-1} displaying
  variability with a 0.8\,mag amplitude. The object is also listed in
  outburst in the USNO-B1.0 catalogue.  The spectrum is consistent
  with that of a blazar, but the object does not have a radio
  counterpart.  The untypical spectrum warrants additional
  observations.

\item \textit{SDSS\,J170145.85+332339.5.} The Gemini spectrum is
  typical of a quiescent dwarf nova (Fig.\,\ref{f-gemini}), with broad
  Balmer and He\,I emission lines.

\item \textit{SDSS\,J170810.31+445450.7.} The ACAM spectrum
  (Fig.\,\ref{f-acam}) contains broad Balmer and He\,I emission lines,
  typical of a dwarf nova in quiescence.

\item \textit{SDSS\,J171202.95+275411.0.}  The Gemini spectrum
  (Fig.\,\ref{f-gemini}) shows strong Balmer and He\,I emission lines,
  typical of a quiescent dwarf nova.

\subsection{Systems without spectra}

\item \textit{SDSS\,J064911.48+102322.1.} The second SDSS observation
  was carried out only two days after the initial and brightest
  detection, so the system may actually have still been in the
  declining phase of the outburst rather than back in quiescence.  In
  fact, the object was not found in other catalogues, suggesting that
  it is substantially fainter in quiescence.

\item \textit{SDSS\,J074500.58+332859.5.} The long-term light curve of
  this object is not consistent with being a dwarf nova. SDSS detected
  the object twice at $r\simeq22.2$ on MJD 51874 and 51962, and once
  at $r=18.9$ on MJD 54138. CRTS monitoring started around MJD 53319,
  but did not detect SDSS\,J0745+3328 until MJD $\simeq53449$, when it
  reached $r\simeq18.5$. The object stayed at that brightness level
  for $\simeq600$ days, displaying short-term variability with a
  peak-to-peak magnitude of $\simeq1$\,mag, and then faded over a
  period of $\simeq400$ days back to the CRTS detection threshold
  $\simeq21$ over a period of $\simeq400$ days. Both the long duration
  of the high and low states, as well as the observed magnitude
  difference between the two states are reminiscent of the long-term
  light curves of polars, and we speculate that SDSS\,J0745+3328 is a
  polar.  Because it was not detected at radio wavelengths, it is not
  likely to be a blazar.  SDSS\,J0745+3328 was not included in the 
  further statistical analysis in Section\,\ref{s-discussion}.

\item \textit{SDSS\,J105333.76+285033.6.} The CRTS light curves
  exhibits extremely frequent brightness variations, suggesting very
  frequent outbursts, or that the system is eclipsing.

\item \textit{SDSS\,J132040.96-030016.7.} This system was found
  initially from cross-matches with GALEX by relaxing the
  conditions~(\ref{colourcond}) and instead imposing $|fuv
  - nuv| < 1$ in addition to a large UV excess.  All the SDSS colours
  fit the conditions~(\ref{colourcond}), except $g-r = 4.02$.  As the
  sequence of filters in the SDSS photometry is $r$, $i$, $u$, $z$,
  $g$, it is possible that the white dwarf and/or the accretion disk
  was eclipsed by the red dwarf during this time. The source is also
  detected in the ROSAT All Sky Survey Bright Source Catalogue
  \citep{vogesetal99-1}, making it particularly interesting for
  follow-up observations.

\item \textit{SDSS\,J212025.17+194156.3.} This object was also
  detected in outburst by CRTS during the preparation of this paper
  (August 2009).

\end{list}

%
%______________________________________________________________

\section{Discussion}
\label{s-discussion}

For the remainder of the paper we will work with all of the new dwarf novae found above as a single sample,
irrespective of the method they were found with, as the individual subsamples are too small.
As indicated above a bias toward short period CVs was introduced by method ``D'', and this will be carried
forward, although in a diminished form, into the complete sample.

\begin{figure}
\centering
\includegraphics[width=\columnwidth]{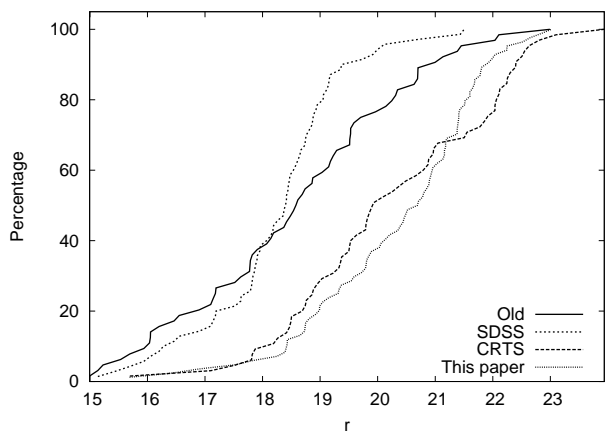}
\caption{ Cumulative distribution of the dwarf novae with
  SDSS data, according to their $r$ magnitude in quiescence.}
\label{Distribution}
\end{figure}

Table~\ref{Statistics} gives an overview of dwarf novae which have
SDSS photometric data at minimum, divided into bins of one magnitude
centred on the given $r$ value in the first column.  The second
column in the table gives the number of dwarf novae from our original
sample (Table~\ref{DNSample}, excluding 19 objects which were only
observed in outburst by SDSS).  The third column gives the number of
those systems that were \textit{not} discovered by SDSS or CRTS, but
were identified by other methods (mostly variability), and we will
denote this sub-sample as ``old'' in the discussion that follows.  The
fourth column gives the number of dwarf novae from
Table~\ref{DNSample} for which SDSS-obtained spectroscopy has been
published by \citet[][and previous papers in that
  series]{szkodyetal09-1}, irrespective of whether they were already
known before, the fifth column those that were detected in outburst by
CRTS (including objects from the ``old'' list), and the sixth column
the number of known systems recovered in this study.  The last
column lists the number of new dwarf nova candidates found in this
paper, excluding 9 with only SDSS data in outburst.
Fig.\,\ref{Distribution} illustrates the normalised cumulative
distribution of the quiescent magnitudes of the dwarf novae not
discovered by SDSS or CRTS (``old'' systems), those detected
spectroscopically by SDSS, those found in outburst by CRTS, and those
detected in this paper (both already known systems as well as new
ones).  The number of new systems from CRTS and our study increases
dramatically compared to the previously known ones for $r\ga21$.  A
Kolmogorov-Smirnov test shows that there is less than 1\%
probability that the magnitude distribution of the CRTS and our new
dwarf novae differ from that of the previously known ones.  From the fact
  that about as many dwarf novae were recovered as new ones found for
  $r \le 20$ (21 compared to 19) one may conclude that about half of
  the regularly outbursting dwarf novae are known with $r \le 20$
  (including recent CRTS discoveries) in the SDSS fields.  Outside the
  SDSS fields there will be a much smaller fraction of the systems
  known, as SDSS contributed significantly to the (spectroscopic)
  identification of dwarf novae.  Also, only a small fraction of fainter systems
  is known.  These may be objects that are simply further away, or
may belong to a population of intrinsically fainter objects with lower
mass transfer rates.

The apparently fainter systems were
predominantly discovered by CRTS, as the SDSS spectroscopy does not
reach deep enough.
Only experiments like CRTS, which are similar in nature to
the way the study presented here was set up, are able to
discover them at this time.
%Although the CRTS distribution shows a
%dip near $r=21$ in Fig.~\ref{Distribution}, it is not significantly
%different from the distribution of dwarf nova detected in this paper.
%The dip seems to be an artefact of sparse data, as it is not present
%in the distribution of the minimum brightness of all dwarf novae
%detected by CRTS (see further and Fig.~\ref{CRTSDist}).
The number of dwarf novae found in this paper decreases significantly
for $r > 21$ at quiescence, essentially because of the magnitude limit
of SDSS and the outburst cutoff at 18th magnitude used for
matching with the astrometric catalogues.  CRTS goes deeper than this
and still discovers a large number of systems with $r \approx 22$.

\begin{table}
\begin{center}
\caption{Number of dwarf novae binned according to their $r$
    magnitude in quiescence, based on the sample from
  Table~\ref{Statistics} and the dwarf nova candidates in
  Table~\ref{CVs}.  See text for details.}  \vskip2mm
\label{Statistics}
\begin{tabular}{crrrrrr}
\hline
$r$ & Known & Old & SDSS & CRTS & Recovered & New \\
\hline
15 & 4 & 3 & 2 & 0 & 0 & 0 \\
16 & 9 & 8 & 6 & 1 & 2 & 0 \\
17 & 6 & 6 & 6 & 1 & 1 & 0 \\
18 & 31 & 13 & 27 & 9 & 3 & 4 \\
19 & 29 & 12 & 22 & 13 & 9 & 4 \\
20 & 26 & 11 & 4 & 13 & 6 & 11 \\
21 & 17 & 8 & 2 & 7 & 6 & 19 \\
22 & 15 & 2 & 1 & 15 & 1 & 14 \\
23 & 6 & 1 & 0 & 5 & 1 & 3 \\
24 & 1 & 0 & 0 & 1 & 0 & 0 \\
\hline
Total & 144 & 64 & 69 & 65 & 29 & 55 \\
\hline

\end{tabular}
\end{center}
\end{table}

\begin{figure}
\centering
\includegraphics[width=\columnwidth]{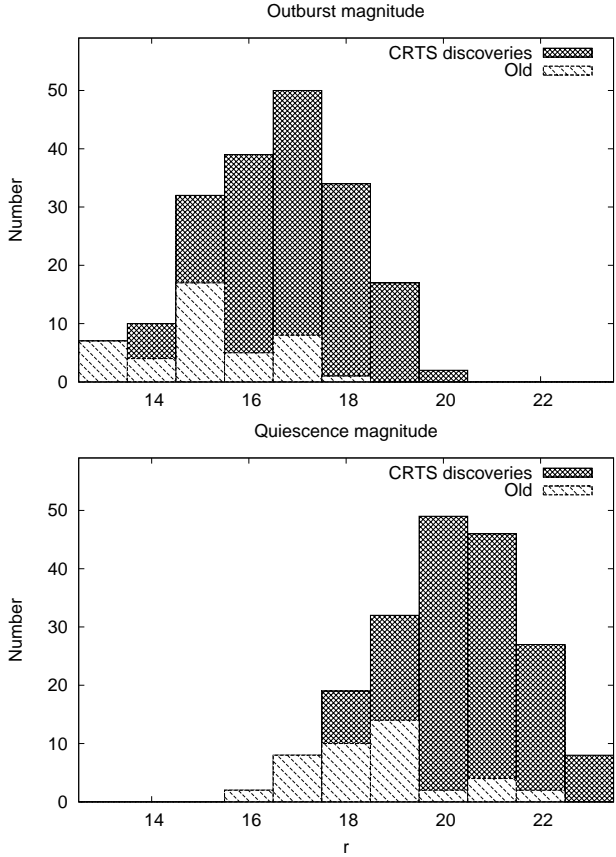}
\caption{Distribution of previously known and newly discovered dwarf
  novae detected in outburst by CRTS.}
\label{CRTSDist}
\end{figure}

The fact that CRTS is discovering mostly faint systems is
clearly illustrated in Fig.~\ref{CRTSDist}.  The top panel
compares the distribution of the observed outburst magnitudes
  of the dwarf novae found by CRTS and those known before, whereas the
  bottom panel shows the same for the quiescent magnitudes.
Note that a minimum magnitude is often difficult to obtain for
  CRTS objects fainter than magnitude 22, at which level magnitudes
  can no longer be trusted. For a number of systems no reliable quiescent
  magnitude could be obtained, in which case it was set to a default
  of 23. Consequently, the outburst amplitudes may be underestimated
  in these cases.  The data are based on CRTS outbursts for 42
previously known dwarf novae and 149 newly discovered systems from the
whole sky observed by CRTS (not limited to SDSS fields) in the period
of November 2007 (when CRTS operations started) to May 2009.  CRTS
detections are not biased towards either new or already known dwarf
novae, both are reported in the same way.  CRTS did not discover any new
dwarf novae with $r \leq 17$ at quiescence.
%So it is likely that almost all of the
%brighter dwarf novae with frequent outbursts are known.  For outburst
%magnitudes $r \geq 16$ and quiescent magnitudes $r
%\geq 19$, more than half of the systems detected in outburst by CRTS
%were not known before.

\begin{figure}
\centering
\includegraphics[width=\columnwidth]{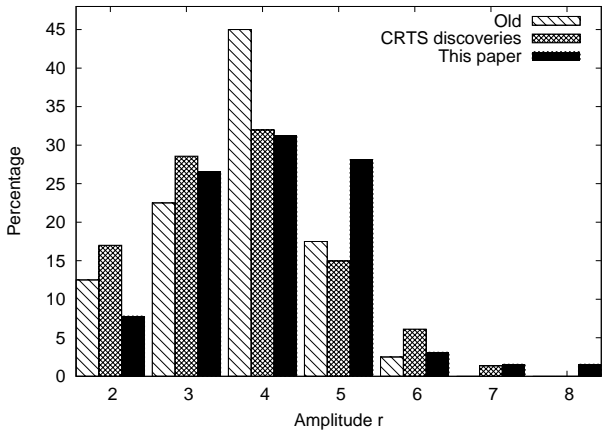}
\caption{Distribution of the outburst amplitudes of the
  previously known and newly discovered dwarf novae detected in
  outburst by CRTS and of the new systems found in this
  paper. }
\label{CRTSAmp}
\end{figure}

The outburst amplitude may be a characteristic that reflects differences
between ``normal'' and intrinsically faint dwarf novae,
as WZ~Sge systems only show large amplitude superoutbursts.
Although the largest outburst amplitude dwarf novae tend to
be new CRTS discoveries, the distribution of the amplitudes between
previously known and CRTS discovered systems are very much alike
(Fig.~\ref{CRTSAmp}).
The new dwarf novae presented in this paper also tend to have larger
amplitudes.  This is in part due to a selection effect because the
constraints used to detect them, favour those systems with larger
amplitudes.  However, as for the CRTS-discovered dwarf
  novae the tail of the distribution is also longer than for the
previously known systems.

Lacking orbital period information, a difference in dwarf novae
populations can also be deduced from a difference in outburst
frequency.  Our sample of observed outbursts is, however, too
fragmentary compared to those of the known systems to draw any
firm conclusions.  In order to get a better estimate for the
outburst frequency, CRTS data of known systems\footnote{Available from
  http://nesssi.cacr.caltech.edu/catalina/CVservice/CVtable.html} were
studied and compared to the CRTS data of the new objects.  Only light
curves with more than 20 data points were considered (i.e. effectively
observed on at least five nights).  The available data had a timespan
from three to almost five years, and up to 292 data points were
available.  A dwarf nova was considered to be in outburst if its
brightness differed by at least two magnitudes from its
  quiescent value.  The number of data points in outburst divided by
the total number of observations is then a measure of the outburst
frequency.  Table~\ref{Freq} lists the outburst frequencies
  determined in this way for the previously known systems, excluding
CRTS and SDSS discoveries (second column), SDSS discovered dwarf novae
(third column), CRTS discoveries (fourth column) and the new systems
from Table~\ref{CVs} (last column).  CRTS discoveries are biased
because they had to be in outburst for them to be discovered.  But
considering only dwarf novae which are observed at least once in
outburst by CRTS, the major difference seems to be the large
proportion of known systems which are active a large fraction of the
time (mostly Z~Cam type systems).
% and the fact that the new discoveries (including CRTS) tend to have the
% smallest outburst frequency.

The previously known systems were more likely to be found in outburst
than the new ones.
This is the reverse of what is expected as all new systems were found by outbursts,
while some of the old ones were not.
Considering only the fact whether new and old
systems were found to be in outburst by CRTS or not,
 Fisher's exact test gives
only a 3\% probability that the same or a more extreme
difference is observed for samples of the same population.
Still, it is not
possible to conclude that the new and old populations are entirely
different on the basis of this statistic alone, but there is more than a hint that on average, the
new systems are not only fainter, but also less active.
This supports
the conclusion from \citet{gaensickeetal09-2} that deep CV
  samples contain a significant fraction of intrinsically faint
  systems with low accretion rates.  At first sight there
seems to be no difference in the outburst frequency of the previously
known dwarf novae and those discovered by SDSS spectroscopically
(e.g. the fraction of systems without outbursts is essentially the
same), which may seem a contradiction.  However note that the the SDSS
sample considered in this paper is only a subset of all CVs found by
SDSS, namely those which have had at least one observed outburst.  It
is very likely that among the remainder of the SDSS CVs there is
a significant number of infrequently outbursting systems.

\begin{table}
\begin{center}
\caption{Estimated outburst frequency for confirmed dwarf novae
  derived from CRTS observations.} \vskip2mm
\label{Freq}
\begin{tabular}{rcrrrrr}
\hline
\multicolumn{3}{c}{Time in} & \multicolumn{1}{c}{Previously} & \multicolumn{1}{c}{SDSS} & \multicolumn{1}{c}{CRTS} & \multicolumn{1}{c}{This}  \\
\multicolumn{3}{c}{outburst (\%)} & \multicolumn{1}{c}{known} & \multicolumn{1}{c}{discoveries} & \multicolumn{1}{c}{discoveries} & \multicolumn{1}{c}{paper} \\
\hline
\multicolumn{3}{c}{No outburst} & 38 & 10 & 0 & 25 \\
0 & - & 5 & 21 & 10 & 33 & 7 \\
5 & - & 10 & 22 & 7 & 73 & 7 \\
10 & - & 15 & 16 & 6 & 24 & 4 \\
15 & - & 20 & 7 & 1 & 15 & 5 \\
20 & - & 25 & 3 & 0 & 5 & 0 \\
25 & - & 30 & 6 & 1 & 2 & 4 \\
30 & - & 35 & 0 & 1 & 0 & 1 \\
35 & - & 40 & 0 & 2 & 2 & 0 \\
40 & - & 50 & 3 & 0 & 2 & 0 \\
50 & - & 100 & 10 & 1 & 1 & 1 \\
\hline
\multicolumn{3}{c}{Total} & 126 & 39 & 157 & 54 \\
\hline
\end{tabular}
\end{center}
\end{table}

The time CRTS has been in operation is too short to study changes in
the rate at which it is discovering new dwarf novae.  This could have
given an indication of how many systems are still to be discovered.
But both the CRTS discoveries and the new dwarf novae found in this
paper suggest a large population of faint dwarf novae of which only a
small fraction is known at this time.  Detailed studies of
  this kind will become more feasible in the intermediate future when
  more CRTS observations are available, and in the longer term through
  LSST.

%
%______________________________________________________________

\section{Conclusion}

In this paper 64 new CV candidates, mostly dwarf novae, are presented,
resulting from data mining between existing photometric and
astrometric catalogues. 16 of these systems are unmistakably
  confirmed as CVs through follow-up spectroscopy.  It is shown that
a large number of faint CVs are missed by surveys like SDSS, but can
be picked up by either transient surveys like CRTS, or by archival
plate studies.  The importance of having historical plates available
electronically and measured, cannot be overestimated in this regard.
Even including the recent CRTS discoveries, it is estimated that fewer
than half the dwarf novae with a quiescent $r \le 20$
are currently known in the SDSS fields, and substantially fewer of the fainter
ones.  It appears that CVs will constitute a significant source of
transients for future surveys such as LSST and Pan-STARRS.

Studying amplitude and outburst frequency suggests that the new faint
dwarf novae consist of systems both similar to the previously known
ones but simply farther away, and of intrinsically fainter systems.
The data that will become available in the near future from CRTS and
the planned surveys, will give an opportunity to study the latter
population in much more detail.

Both SDSS\,J093946.03+065209.4 and SDSS\,J162900.55 +341022.0, two of
the dwarf novae candidates found, have unusual spectra for cataclysmic
variables and deserve to be studied in more detail.

%
%______________________________________________________________

\section*{Acknowledgements}

This study made use of Simbad and VizieR \citep{ochsenbeinetal00-1}, the USNOFS Image and
Catalogue Archive operated by the United States Naval Observatory,
Flagstaff Station, optical
images generated by the Near Earth Asteroid Tracking (NEAT) through
the Skymorph website and of data
provided by the GALEX mission and the Sloan Digital Sky Survey (SDSS).
GALEX (Galaxy Evolution Explorer) is a NASA Small Explorer, launched
in April 2003.  It is operated for NASA by Caltech under NASA contract
NAS5-98034.
%http://cas.sdss.org/astro/en/credits/datause.asp
Funding for the SDSS has been provided by the Alfred P. Sloan
Foundation, the Participating Institutions, the National Aeronautics
and Space Administration, the National Science Foundation, the
U.S. Department of Energy, the Japanese Monbukagakusho, and the Max
Planck Society. The SDSS Web site is http://www.sdss.org/.  The SDSS
is managed by the Astrophysical Research Consortium (ARC) for the
Participating Institutions.  The Participating Institutions are The
University of Chicago, Fermilab, the Institute for Advanced Study, the
Japan Participation Group, The Johns Hopkins University, Los Alamos
National Laboratory, the Max-Planck-Institute for Astronomy (MPIA),
the Max-Planck-Institute for Astrophysics (MPA), New Mexico State
University, University of Pittsburgh, Princeton University, the United
States Naval Observatory, and the University of Washington.

The CSS survey is funded by the National Aeronautics and Space Administration
under Grant No. NNG05GF22G issued through the Science Mission Directorate Near-Earth Objects Observations Program.
The CRTS survey is supported by the U.S. National Science Foundation under grant No. AST-0909182.

This paper is based in part on observations made with the William Herschel Telescope
operated on the island of La Palma by the Isaac Newton Group in the Spanish Observatorio del Roque de los Muchachos
of the Instituto de Astrof\'isica de Canarias and on observations (GN-2008A-Q-86 and GS-2009A-Q-73)
obtained at the Gemini Observatory, which is operated by the Association of Universities for Research in Astronomy, Inc.,
under a cooperative agreement with the NSF on behalf of the Gemini partnership: the National Science Foundation (United States),
the Science and Technology Facilities Council (United Kingdom), the National Research Council (Canada), CONICYT (Chile),
the Australian Research Council (Australia), Minist\'erio da Ci\^encia e Tecnologia (Brazil)
and Ministerio de Ciencia, Tecnología e Innovaci\'on Productiva (Argentina).

We thank Areg Mickaelian for providing us with copies of the paper by \citet{hovanissian78-1,hovanissian82-1}.
We also thank the referee, Christian Knigge, for his constructive remarks to improve the paper.

%\bibliographystyle{mn_new}
%\bibliography{aamnem99,aabib,proceedings,submitted}

\begin{thebibliography}{31}
\expandafter\ifx\csname natexlab\endcsname\relax\def\natexlab#1{#1}\fi

\bibitem[{{Abazajian} et~al.(2005)}]{abazajian05-1}
{Abazajian}, K., et~al., 2005, AJ, 129, 1755

\bibitem[{{Abazajian} et~al.(2009)}]{abazajian09-1}
{Abazajian}, K.~N., et~al., 2009, ApJS, 182, 543

\bibitem[{{Abraham} \& {Carrara}(1998)}]{abraham+carrara98-1}
{Abraham}, Z., {Carrara}, E.A., 1998, ApJ, 496, 172

\bibitem[{{Aungwerojwit} et~al.(2006)}]{aungwerojwit06-1}
{Aungwerojwit}, A., et~al., 2006, A\&A, 455, 659

\bibitem[{{Budav{\'a}ri} et~al.(2009)}]{budavarietal09-1}
{Budav{\'a}ri}, T., et~al., 2009, ApJ, 694, 1281

\bibitem[{{Copenhagen University Obs.} et~al.(2006){Copenhagen University
  Obs.}, {Institute of Astronomy, Cambridge}, \& {Real Instituto y Observatorio
  de la Armada en San Fernando}}]{cmc14}
{Copenhagen University Obs.}, {Institute of Astronomy, Cambridge}, U., {Real
  Instituto y Observatorio de la Armada en San Fernando}, 2006, Carlsberg
  Meridian Catalog Number 14

\bibitem[{{Drake} et~al.(2009)}]{drakeetal09-1}
{Drake}, A.~J., et~al., 2009, ApJ, 696, 870

\bibitem[{{Ducourant} et~al.(2006)}]{ducourantetal06-1}
{Ducourant}, C., et~al., 2006, A\&A, 448, 1235

\bibitem[{{Eisenstein} et~al.(2006)}]{eisensteinetal06-1}
{Eisenstein}, D.~J., et~al., 2006, ApJS, 167, 40

\bibitem[{{G{\"a}nsicke}(2005)}]{gaensicke05-1}
{G{\"a}nsicke}, B.~T., 2005, in {Hameury}, J.-M., {Lasota}, J.-P., eds., The
  Astrophysics of Cataclysmic Variables and Related Objects, ASP Conf. Ser.
  330, p.~3

\bibitem[{{G{\"a}nsicke} et~al.(2009)}]{gaensickeetal09-2}
{G{\"a}nsicke}, B.~T., et~al., 2009, MNRAS, 397, 2170

\bibitem[{{Horne}(1986)}]{horne86-1}
{Horne}, K., 1986, PASP, 98, 609

\bibitem[{{Hovanissian}(1978)}]{hovanissian78-1}
{Hovanissian}, E.~Y., 1978, Soobshcheniya. Byurakan. Astrof. Obs., 50, 5

\bibitem[{{Hovanissian}(1982)}]{hovanissian82-1}
{Hovanissian}, E.~Y., 1982, Soobshcheniya. Byurakan. Astrof. Obs., 53, 77

\bibitem[{{Howell} et~al.(2001){Howell}, {Nelson}, \&
  {Rappaport}}]{howelletal01-1}
{Howell}, S.~B., {Nelson}, L.~A., {Rappaport}, S., 2001, ApJ, 550, 897

\bibitem[{{Jones} et~al.(2005){Jones}, {Saunders}, {Read}, \&
  {Colless}}]{jonesetal05-1}
{Jones}, D.~H., {Saunders}, W., {Read}, M., {Colless}, M. et~al., 2005, MNRAS, 355, 747

\bibitem[{{Jones} et~al.(2009){Jones}, {Read}, \&
 {Saunders}}]{jonesetal09-1}
{Jones}, D.~H., {Read}, M.~A., {Saunders} W. et~al., 2009, MNRAS, 399, 683

\bibitem[{{Kolb}(1993)}]{kolb93-1}
{Kolb}, U., 1993, A\&A, 271, 149

\bibitem[{{Kolb} \& {Baraffe}(1999)}]{kolb+baraffe99-1}
{Kolb}, U., {Baraffe}, I., 1999, MNRAS, 309, 1034

\bibitem[{{Lasker} et~al.(2008)}]{laskeretal08-1}
{Lasker}, B.~M., et~al., 2008, AJ, 136, 735

\bibitem[{{Lasota}(2001)}]{lasota01-1}
{Lasota}, J.-P., 2001, New Astronomy Review, 45, 449

\bibitem[{{Lawrence} et~al.(2007)}]{lawrenceetal07-1}
{Lawrence}, A., et~al., 2007, MNRAS, 379, 1599

\bibitem[{{Li} \& {Thakar}(2008)}]{li+thakar08-1}
{Li}, N., {Thakar}, A.~R., 2008, Computing in Science and Engineering, 10, 18

\bibitem[{{Lohmann}(1949)}]{lohmann49-1}
{Lohmann}, W., 1949, Astronomische Nachrichten, 277, 37

\bibitem[{{Marsh}(1989)}]{marsh89-1}
{Marsh}, T.~R., 1989, PASP, 101, 1032

\bibitem[{{Martin} et~al.(2005)}]{martinetal05-1}
{Martin}, D.~C., et~al., 2005, ApJ Lett., 619, L1

\bibitem[{{Meyer} \& {Meyer-Hofmeister}(1984)}]{meyer+meyer-hofmeister84-1}
{Meyer}, F., {Meyer-Hofmeister}, E., 1984, A\&A, 132, 143

\bibitem[{{Monet} et~al.(2003)}]{monetetal03-1}
{Monet}, D.~G., et~al., 2003, AJ, 125, 984

\bibitem[{{Ochsenbein} et~al.(2000){Ochsenbein}, {Bauer}, \&
  {Marcout}}]{ochsenbeinetal00-1}
{Ochsenbein}, F., {Bauer}, P., {Marcout}, J., 2000, A\&AS, 143, 23

\bibitem[{{Osaki}(1996)}]{osaki96-1}
{Osaki}, Y., 1996, PASP, 108, 39

\bibitem[{{Pesch} \& {Sanduleak}(1987)}]{pesch+sanduleak87-1}
{Pesch}, P., {Sanduleak}, N., 1987, Information Bulletin on Variable Stars,
  2989, 1

\bibitem[{{Pretorius} et~al.(2007){Pretorius}, {Knigge} \& {Kolb}}]{pretoriusetal07-1}
{Pretorius}, M.~L., {Knigge}, C., {Kolb}, U., 2007, MNRAS, 374, 1495

\bibitem[{{Pretorius} \& {Knigge}(2008)}]{pretorius+knigge08-1}
{Pretorius}, M.~L., {Knigge}, C., 2008, MNRAS, 385, 1485

\bibitem[{{Richards} et~al.(2002)}]{richardsetal02-1}
{Richards}, G.~T., et~al., 2002, AJ, 123, 2945

\bibitem[{{Ritter} \& {Kolb}(2003)}]{ritter+kolb03-1}
{Ritter}, H., {Kolb}, U., 2003, A\&A, 404, 301

\bibitem[{{Szkody} et~al.(2004)}]{szkodyetal04-1}
{Szkody}, P., et~al., 2004, AJ, 128, 1882

\bibitem[{{Szkody} et~al.(2009a)}]{szkodyetal09-1}
{Szkody}, P., et~al., 2009a, AJ, 137, 4011

\bibitem[{{Szkody} et~al.(2009b)}]{szkodyetal09-2}
{Szkody}, P., et~al., 2009b, in preparation

\bibitem[{{Voges} et~al.(1999)}]{vogesetal99-1}
{Voges}, W., et~al., 1999, A\&A, 349, 389

\bibitem[{{Warren} et~al.(2007)}]{warrenetal07-1}
{Warren}, S.~J., et~al., 2007, MNRAS, 375, 213

\bibitem[{{Welsh} et~al.(2005)}]{welshetal05-1}
{Welsh}, B.~Y., et~al., 2005, AJ, 130, 825

\bibitem[{{Willems} et~al.(2005){Willems}, {Kolb}, {Sandquist}, {Taam}, \&
  {Dubus}}]{willemsetal05-1}
{Willems}, B., {Kolb}, U., {Sandquist}, E.~L., {Taam}, R.~E., {Dubus}, G.,
  2005, ApJ, 635, 1263

\end{thebibliography}

\clearpage
\appendix
\section{Online Tables}

\setcounter{table}{0}
\begin{table*}
\begin{center}
\caption{Known dwarf novae with SDSS $ugriz$ data.  An asterisk after the $r$ magnitude indicates the
  object was in outburst at the time of the SDSS
  photometry.  The dwarf nova subtypes are defined as follows: UG: U Gem without subclassification; 
UGSS: SS Cyg subtype; UGZ: Z Cam subtype; UGSU: SU UMa subtype, UGWZ: WZ Sge subtype; 
E: eclipsing; ZZ: ZZ Ceti type white dwarf pulsations.
The orbital period, taken from the online Ritter \& Kolb catalogue, is given in hours.
  The recovery method refers to the four sub-steps from Fig.\,2 of the paper: A (match with GALEX), B (SDSS variability), C (no USNO counterpart) and D (matching with astrometric catalogues).
 }
\label{DNSampleFull}
\begin{tabular}{lclllrrrrc} 
\hline
Name                      & SDSS             & Type & $P_{orb}$ & \multicolumn{1}{c}{$r$} & $u-g$ & $g-r$ & $r-i$ & $i-z$ & Method \\
\hline
TY Psc                    & J012539.35+322308.6 & UGSU   & 1.64 & 13.05* & 0.37 & -0.08 & -0.28 & -0.34 & D \\
SDSS J013132.39-090122.3  & J013132.39-090122.2 & UG+ZZ  & 1.36 & 18.42  & 0.02 & -0.12 & -0.18 & 0.13 &  \\
SDSS J013701.06-091235.0  & J013701.06-091234.8 & UGSU   & 1.33 & 18.45  & 0.34 & 0.24 & 0.38 & 0.28 &  \\
SDSS J015151.87+140047.2  & J015151.87+140047.2 & UG     & 1.98 & 20.00  & -0.29 & 0.30 & 0.41 & 0.45 &  \\
SDSS J031051.66-075500.3  & J031051.66-075500.3 & UGSU   & 1.61 & 15.74* & 0.26 & -0.27 & -0.16 & -0.22 & D \\
CSS081107:033104+172540   & J033104.44+172540.2 & UG     &      & 19.33  & 0.09 & 0.54 & 0.45 & 0.41 & D \\
SDSS J033449.86-071047.8  & J033449.86-071047.8 & UGSU   & 1.90 & 14.85* & 0.28 & -0.25 & -0.18 & -0.19 & D \\
SDSS J033710.91-065059.4  & J033710.91-065059.4 & UG     &      & 19.72  & 0.10 & -0.18 & -0.25 & -0.17 &  \\
CSS080303:073921+222454   & J073921.17+222453.5 & UG     &      & 22.40  & 0.06 & 0.27 & 0.27 & -0.37 &  \\
SDSS J074531.91+453829.5  & J074531.91+453829.6 & UG+ZZ  & 1.27 & 18.93  & -0.23 & 0.13 & -0.13 & 0.03 &  \\
SDSS J074640.62+173412.8  & J074640.62+173412.8 & UGSU   & 1.57 & 18.37  & 0.06 & -0.20 & -0.15 & -0.05 &  \\
CSS090331:074928+190452   & J074928.01+190452.1 & UG     &      & 20.89  & -0.56 & 0.22 & 0.07 & 0.50 & A \\
U Gem                     & J075505.20+220004.9 & UGSS   & 4.25 & 10.43* & 0.48 & -0.16 & -0.06 & -0.53 &  \\
CSS080406:075648+305805   & J075648.04+305805.0 & UG     &      & 20.96  & -0.13 & 0.10 & -0.08 & -0.09 & D \\
SDSS J080434.20+510349.2  & J080434.20+510349.1 & UGWZ   & 1.42 & 17.86  & 0.17 & -0.02 & -0.15 & -0.08 &  \\
VSX J080714.2+113812      & J080714.25+113812.5 & UGSU   & 1.44 & 20.70  & 0.20 & -0.16 & -0.27 & 0.65 &  \\
PTF09bd                   & J080729.70+153441.9 & UG     &      & 22.04  & 0.03 & 0.24 & 0.64 & -0.13 &  \\
SDSS J080846.19+313106.0  & J080846.19+313105.9 & UGSU   & 4.94 & 18.74  & -0.30 & 0.69 & 0.58 & 0.43 & B \\
CSS080416:080854+355053   & J080853.72+355053.6 & UG     &      & 19.63  & -0.10 & 0.02 & 0.11 & 0.20 &  \\
YZ Cnc                    & J081056.64+280833.1 & UGSU   & 2.08 & 12.76* & 0.89 & 0.01 & -0.18 & -0.55 & D \\
SDSS J081207.63+131824.4  & J081207.63+131824.4 & UGSU   & 1.95 & 19.16  & -0.04 & 0.10 & 0.13 & 0.36 &  \\
SDSS J081321.91+452809.4  & J081321.92+452809.4 & UG     & 6.94 & 17.63  & -0.09 & 0.63 & 0.44 & 0.30 &  \\
CSS080202:081415+080450   & J081414.92+080449.5 & UG     &      & 22.06  & -0.32 & 0.24 & 0.38 & 0.44 &  \\
CSS090224:082124+454135   & J082123.72+454135.2 & UG     &      & 19.79  & -0.14 & 0.31 & 0.32 & 0.25 &  \\
SDSS J082409.73+493124.4  & J082409.72+493124.4 & UGSU   & 1.58 & 18.81  & -0.38 & 0.40 & 0.16 & 0.15 & D \\
CSS080306:082655-000733   & J082654.69-000733.1 & UG     &      & 19.50  & 0.33 & 0.02 & -0.14 & 0.01 &  \\
CSS071231:082822+105344   & J082821.78+105343.9 & UG     &      & 22.12  & -0.27 & 0.22 & -0.09 & 0.17 &  \\
AT Cnc                    & J082836.92+252003.0 & UGZ    & 4.83 & 13.88* & 0.19 & -0.21 & 0.17 & 0.00 &  \\
DE Cnc                    & J083527.23+194530.6 & UG     &      & 17.78  & 1.63 & 0.70 & 0.31 & 0.20 &  \\
CC Cnc                    & J083619.15+212105.3 & UGSU   & 1.76 & 16.55  & 0.41 & 0.21 & 0.08 & 0.06 &  \\
SW UMa                    & J083642.74+532838.0 & UGSU   & 1.36 & 16.90  & 0.04 & -0.03 & -0.11 & 0.15 & D \\
EI UMa                    & J083821.99+483802.1 & UG     & 6.43 & 16.04  & 0.06 & -1.06 & 1.25 & 0.18 &  \\
SDSS J083845.23+491055.5  & J083845.23+491055.3 & UGSU   & 1.67 & 19.38  & -0.36 & 0.20 & 0.03 & 0.27 &  \\
CSS090331:084041+000520   & J084041.39+000520.3 & UG     &      & 20.79  & -0.25 & 0.01 & -0.19 & 0.47 &  \\
EG Cnc                    & J084303.98+275149.6 & UGWZ   & 1.41 & 18.87  & 0.16 & -0.02 & -0.20 & -0.06 &  \\
CSS080309:084358+425037   & J084358.08+425036.7 & UG     &      & 19.89  & -0.32 & 0.03 & 0.18 & 0.39 & D \\
SDSS J084400.10+023919.3  & J084400.10+023919.3 & UG     & 4.97 & 17.96  & -0.12 & 0.38 & 0.38 & 0.26 &  \\
CSS080209:084555+033929   & J084555.07+033929.1 & UG     & 1.43 & 20.87  & -0.13 & 0.00 & -0.15 & -0.24 &  \\
CT Hya                    & J085107.39+030834.3 & UGSU   & 1.56 & 18.86  & -0.07 & -0.07 & 0.01 & 0.24 &  \\
CSS080401:085113+344449   & J085113.44+344448.5 & UG     & 1.92 & 20.08  & 0.15 & 0.08 & 0.25 & 0.61 &  \\
BZ UMa                    & J085344.16+574840.5 & UGSU   & 1.63 & 16.05  & -0.39 & 0.32 & -0.02 & 0.20 &  \\
CSS080506:085409+201339   & J085409.43+201339.0 & UG     &      & 20.49  & -0.62 & 0.43 & 0.17 & 0.23 & D \\
AK Cnc                    & J085521.18+111815.0 & UGSU   & 1.56 & 18.74  & 0.13 & 0.06 & 0.15 & 0.32 &  \\
CSS071112:085823-003729   & J085822.82-003729.4 & UG     &      & 22.04  & 0.31 & 0.27 & 0.12 & 0.39 &  \\
SDSS J090016.56+430118.2  & J090016.55+430118.1 & UG     & 5.02 & 18.18  & -0.24 & 0.70 & 0.68 & 0.49 &  \\
SY Cnc                    & J090103.30+175355.9 & UGZ    & 9.12 & 12.23* & 1.36 & -0.07 & -0.03 & -0.73 &  \\
SDSS J090103.93+480911.1  & J090103.93+480911.1 & UG+E   & 1.87 & 19.11  & -0.09 & 0.12 & 0.10 & 0.17 &  \\
CSS080304:090240+052501   & J090239.70+052500.7 & UGSU   & 1.36 & 23.10  & 0.59 & 0.07 & 0.10 & 0.88 &  \\
SDSS J090628.25+052656.9  & J090628.24+052656.9 & UG     &      & 18.47  & 0.07 & 0.29 & 0.36 & 0.29 &  \\
GUVV J090904.4+091714.4   & J090904.40+091713.1 & UG     &      & 22.11  & -0.26 & 0.21 & 0.43 & 0.39 &  \\
GY Cnc                    & J090950.53+184947.4 & UG+E   & 4.21 & 15.69  & -0.10 & 0.33 & 0.37 & 0.32 & D \\
DI UMa                    & J091216.19+505353.8 & UGSU   & 1.31 & 17.77  & -0.13 & 0.09 & 0.03 & 0.01 &  \\
CSS081107:091454+113402   & J091453.59+113401.6 & UG     &      & 21.05  & 0.25 & -0.08 & -0.15 & -0.10 &  \\
PQT 080119:091534+081356  & J091534.90+081356.0 & UG     &      & 22.99  & -0.54 & -0.20 & 0.32 & 0.92 &  \\
GZ Cnc                    & J091551.67+090049.5 & UG     & 2.11 & 15.53  & 0.65 & -0.80 & 0.68 & 0.17 &  \\
HH Cnc                    & J091650.76+284943.1 & UGSS   &      & 18.54  & -0.17 & 0.63 & 0.34 & 0.26 &  \\
AR Cnc                    & J092207.56+310314.4 & UG+E   & 5.15 & 19.59  & 0.87 & 1.41 & 0.84 & 0.58 & A \\
\hline
\end{tabular}
\end{center}
\end{table*}

\begin{table*}
\begin{center}
\contcaption{}
\begin{tabular}{lclllrrrrc} 
\hline
Name                      & SDSS             & Type & $P_{orb}$ & \multicolumn{1}{c}{$r$} & $u-g$ & $g-r$ & $r-i$ & $i-z$ & Method \\
\hline
Lyn2                      & J093423.39+403808.9 & UG     &      & 19.96  & -0.51 & 0.16 & 0.00 & 0.25 &  \\
HM Leo                    & J093836.98+071455.0 & UG     & 4.48 & 17.79  & -0.05 & 0.53 & 0.72 & 0.61 &  \\
DV UMa                    & J094636.59+444644.7 & UGSU+E & 2.06 & 19.15  & -0.04 & 0.24 & 0.34 & 0.56 & D \\
ER UMa                    & J094711.93+515409.0 & UGSU   & 1.53 & 13.77* & 0.56 & 0.00 & -0.75 & 0.38 &  \\
X Leo                     & J095101.46+115231.2 & UG     & 3.95 & 16.05  & -0.06 & 0.28 & 0.36 & 0.30 &  \\
RZ LMi                    & J095148.93+340723.9 & UGSU   & 1.40 & 14.81* & 0.38 & -0.22 & -0.19 & -0.20 &  \\
RU LMi                    & J100207.45+335100.2 & UG     & 6.02 & 15.48* & 0.29 & -0.15 & -0.15 & -0.14 &  \\
SDSS J100515.38+191107.9  & J100515.38+191107.9 & UGSU   & 1.80 & 18.26  & -0.06 & -0.06 & 0.13 & 0.29 & D \\
ROTSE3 J100932.2-020155   & J100932.15-020154.5 & UGSU   &      & 20.69  & -0.02 & -0.20 & -0.36 & 0.01 &  \\
HS 1016+3412              & J101947.26+335753.6 & UG     & 1.91 & 18.19  & -0.55 & 0.19 & 0.13 & 0.16 &  \\
KS UMa                    & J102026.52+530433.1 & UGSU   & 1.63 & 17.19  & -0.14 & 0.23 & 0.02 & 0.22 &  \\
Var Leo 2006              & J102146.44+234926.3 & UGWZ   & 1.33 & 20.63  & 0.09 & 0.11 & -0.21 & 0.40 &  \\
NSV 4838                  & J102320.27+440509.8 & UGSU   & 1.63 & 18.69  & -0.32 & 0.14 & 0.12 & 0.06 &  \\
SDSS J102637.04+475426.3  & J102637.04+475426.3 & UGSU   & 1.58 & 20.12  & 0.10 & -0.03 & -0.09 & 0.35 &  \\
CSS080305:102938+414046   & J102937.74+414046.3 & UG     &      & 22.30  & -0.11 & -0.04 & -0.79 & 1.09 &  \\
SDSS J103147.99+085224.3  & J103147.98+085224.3 & UG     &      & 18.76  & 0.00 & 0.05 & 0.20 & 0.26 & D \\
CSS080208:103317+072119   & J103317.27+072118.5 & UG     &      & 19.82  & 0.03 & 0.11 & 0.06 & 0.27 &  \\
SS LMi                    & J103405.42+310808.2 & UGWZ   &      & 19.53  & 1.71 & 1.40 & 0.80 & 0.44 &  \\
IY UMa                    & J104356.72+580731.9 & UGSU   & 1.77 & 17.62  & 0.21 & -0.11 & 0.01 & 0.43 &  \\
SX LMi                    & J105430.43+300610.2 & UGSU   & 1.61 & 16.46  & -0.48 & 0.24 & 0.10 & -0.05 &  \\
EL UMa                    & J105507.00+365945.6 & UG     &      & 20.35  & 0.12 & -0.07 & -0.29 & -0.03 & D \\
CSS080130:105550+095621   & J105550.08+095620.4 & UG     &      & 18.49  & -0.12 & 0.65 & 0.61 & 0.47 &  \\
CY UMa                    & J105656.99+494118.2 & UGSU   & 1.67 & 17.52  & -0.28 & 0.26 & 0.17 & 0.16 & D \\
HS 1055+0939              & J105756.29+092314.9 & UGSS   & 9.02 & 15.23  & 0.65 & 0.68 & 0.29 & 0.16 &  \\
CSS081025:105835+054706   & J105835.11+054706.2 & UG     &      & 20.25  & -0.31 & 0.16 & 0.10 & 0.22 &  \\
2QZ J112555.7-001639      & J112555.71-001638.5 & UG     & 1.51 & 19.52  & -0.15 & 0.12 & -0.11 & 0.36 &  \\
MR UMa                    & J113122.39+432238.5 & UGSU   & 1.53 & 15.94  & -0.33 & 0.20 & 0.05 & -0.10 & C \\
SDSS J113551.09+532246.2  & J113551.09+532246.2 & UG     &      & 20.71  & -0.06 & 0.09 & 0.34 & 0.34 & D \\
RZ Leo                    & J113722.24+014858.5 & UGWZ   & 1.82 & 18.58  & -0.10 & 0.13 & 0.39 & 0.57 &  \\
QZ Vir                    & J113826.82+032207.1 & UGSU   & 1.41 & 15.14  & -0.18 & -0.28 & 0.08 & 0.24 &  \\
NSV 5285                  & J113950.58+455817.9 & UGSU   & 2.03 & 19.51  & -0.01 & 0.12 & 0.13 & 0.49 &  \\
BC UMa                    & J115215.82+491441.8 & UGSU   & 1.50 & 18.61  & 0.12 & -0.09 & -0.01 & 0.24 &  \\
CSS080201:115330+315836   & J115330.25+315835.9 & UG     &      & 19.93  & 0.54 & 0.19 & 0.06 & -0.07 &  \\
SDSS J120231.01+450349.1  & J120231.00+450349.1 & UG     &      & 19.92  & -0.02 & 0.05 & -0.20 & 0.18 &  \\
2QZ J121005.3-025543      & J121005.33-025543.8 & UG     &      & 21.00  & 0.23 & -0.14 & -0.28 & 0.67 &  \\
SDSS J122740.82+513924.9  & J122740.82+513924.9 & UGSU+E & 1.51 & 19.06  & 0.06 & 0.04 & 0.01 & 0.38 &  \\
FU Com                    & J123051.47+270709.8 & UG     &      & 18.48  & 1.20 & 0.37 & 0.14 & 0.06 &  \\
AL Com                    & J123225.79+142042.2 & UGWZ   & 1.36 & 15.46* & 0.21 & -0.21 & -0.14 & -0.10 & D \\
SDSS J123813.73-033933.7  & J123813.73-033932.9 & UGWZ   & 1.34 & 17.82* & 0.11 & -0.04 & -0.16 & -0.10 &  \\
IR Com                    & J123932.00+210806.2 & UGSU+E & 2.09 & 18.37  & -0.07 & -0.05 & 0.36 & 0.78 &  \\
Vir5                      & J124325.92+025547.3 & UG     &      & 18.07  & -0.31 & 0.23 & -0.01 & 0.11 &  \\
CSS080427:124418+300401   & J124417.89+300401.0 & UG     &      & 18.50  & -0.03 & 0.11 & 0.11 & 0.27 &  \\
SDSS J124819.36+072049.4  & J124819.36+072049.4 & UG     &      & 21.50  & 0.35 & -0.14 & 0.18 & -0.26 &  \\
SDSS J125023.85+665525.5  & J125023.84+665525.4 & UG+E   & 1.41 & 18.67  & 0.08 & 0.01 & -0.07 & 0.07 &  \\
GO Com                    & J125637.11+263643.2 & UGSU   & 1.58 & 17.91  & 0.03 & 0.06 & 0.02 & 0.19 &  \\
CSS080702:130030+115101   & J130030.33+115101.2 & UGSU   & 1.51 & 19.80  & 0.10 & 0.01 & -0.05 & 0.44 &  \\
2QZ J130441.7+010330      & J130441.76+010330.8 & UG     &      & 20.24  & -0.17 & -0.09 & -0.03 & 0.31 &  \\
HV Vir                    & J132103.16+015329.0 & UGWZ   & 1.37 & 19.23  & 0.30 & -0.04 & -0.13 & -0.03 &  \\
CSS090102:132536+210037   & J132536.05+210036.7 & UG     &      & 20.37  & -1.17 & 2.73 & -0.25 & 0.09 &  \\
HS 1340+1524              & J134323.16+150916.8 & UG     & 1.54 & 17.14  & -0.34 & 0.18 & 0.04 & 0.04 &  \\
SEKBO 106646.2532         & J140453.98-102702.1 & UG     &      & 19.85  & 0.28 & -0.11 & -0.03 & 0.31 & D \\
OU Vir                    & J143500.22-004606.3 & UGSU   & 1.75 & 18.42  & -0.12 & 0.14 & 0.02 & 0.06 &  \\
RX J1437.0+2342           & J143703.37+234227.9 & UG     &      & 20.31  & 0.30 & -0.15 & -0.25 & -0.07 &  \\
UZ Boo                    & J144401.20+220054.7 & UGWZ   & 1.45 & 19.71  & 0.30 & -0.01 & -0.20 & -0.08 &  \\
TT Boo                    & J145744.75+404340.6 & UGSU   & 1.81 & 19.29  & 0.01 & 0.01 & 0.15 & 0.38 & A \\
EQ 1502+09                & J150441.76+084752.6 & UG     &      & 19.18  & -0.50 & -0.06 & 0.07 & 0.35 &  \\
CSS080514:151021+182303   & J151020.77+182302.0 & UG     &      & 20.66  & 0.11 & 0.73 & 0.18 & 0.23 &  \\
NY Ser                    & J151302.29+231508.4 & UGSU   & 2.35 & 16.20  & 0.17 & -0.12 & -0.06 & -0.01 &  \\
ROTSE3 J151453.6+020934.2 & J151453.63+020934.5 & UG     &      & 15.75* & 0.12 & -0.30 & -0.22 & -0.21 & D \\
SDSS J152212.20+080340.9  & J152212.20+080340.9 & UG     &      & 18.44  & -0.10 & -0.06 & 0.02 & 0.15 &  \\
SDSS J152419.33+220920.0  & J152419.33+220920.0 & UGSU+E & 1.57 & 18.88  & -0.01 & 0.15 & 0.09 & 0.29 &  \\
\hline
\end{tabular}
\end{center}
\end{table*}

\begin{table*}
\begin{center}
\contcaption{}
\begin{tabular}{lclllrrrrc} 
\hline
Name                      & SDSS             & Type & $P_{orb}$ & \multicolumn{1}{c}{$r$} & $u-g$ & $g-r$ & $r-i$ & $i-z$ & Method \\
\hline
QW Ser                    & J152613.96+081802.3 & UGSU   & 1.79 & 17.82  & -0.01 & -0.03 & 0.13 & 0.29 & A \\
SDSS J152857.86+034911.7  & J152857.86+034911.7 & UG     &      & 19.35  & -0.43 & 0.17 & 0.18 & 0.06 & B \\
SDSS J153015.04+094946.3  & J153015.04+094946.3 & UG     &      & 18.48  & -0.48 & 0.41 & 0.01 & -0.06 &  \\
CSS080401:153151+152447   & J153150.87+152446.3 & UG     &      & 22.64  & -0.28 & 0.50 & 0.18 & 0.36 & D \\
DM Dra                    & J153412.16+594831.9 & UGSU   & 1.76 & 20.14  & 0.07 & 0.21 & 0.21 & 0.42 &  \\
SDSS J153634.42+332851.9  & J153634.42+332851.9 & UG     &      & 18.95  & -0.23 & 0.26 & 0.16 & 0.18 &  \\
CSS090322:154428+335725   & J154428.10+335726.3 & UG     &      & 21.85  & 0.22 & 0.22 & -0.41 & 0.67 &  \\
CSS080424:155326+114437   & J155325.70+114437.1 & UG     &      & 23.94  & -0.03 & -0.67 & 0.97 & -0.21 &  \\
CSS081009:155431+365043   & J155430.55+365042.0 & UG     & 1.66 & 21.55  & 0.02 & 0.17 & 0.04 & -0.61 &  \\
SDSS J155644.24-000950.2  & J155644.22-000950.2 & UGSU   & 1.92 & 17.90  & 0.27 & 0.12 & 0.23 & 0.39 &  \\
QZ Ser                    & J155654.47+210718.9 & UG     & 2.00 & 17.18  & 0.57 & 0.73 & 0.32 & 0.22 &  \\
VW CrB                    & J160003.69+331114.1 & UGSU   & 1.70 & 19.53  & 0.11 & 0.11 & 0.08 & 0.14 & A \\
CSS080331:160205+031632   & J160204.80+031631.9 & UG     &      & 22.81  & -0.13 & -0.18 & 0.16 & 1.53 &  \\
CSS080424:160232+161732   & J160232.14+161731.7 & UG     &      & 21.78  & 0.61 & 0.13 & -0.04 & 0.53 &  \\
CSS080428:160524+060816   & J160524.22+060815.9 & UG     &      & 22.51  & -0.26 & 0.34 & 0.48 & -0.54 &  \\
CSS080302:160845+220610   & J160844.81+220609.5 & UG     &      & 20.97  & -0.51 & 0.69 & -0.01 & 0.10 &  \\
SDSS J161332.56-000331.0  & J161332.55-000331.0 & UG     &      & 17.96  & -0.02 & 0.66 & 0.49 & 0.32 &  \\
SDSS J161909.10+135145.5  & J161909.10+135145.5 & UG     &      & 17.81  & 0.45 & 0.68 & 0.39 & 0.27 &  \\
CSS080415:162012+115257   & J162012.06+115256.5 & UG     &      & 22.26  & 0.28 & -0.03 & 0.03 & -0.57 &  \\
V589 Her                  & J162207.15+192236.6 & UGSU   &      & 18.14  & -0.22 & 0.47 & -0.03 & 0.20 &  \\
V844 Her                  & J162501.74+390926.3 & UGSU   & 1.31 & 17.10  & -0.08 & 0.09 & 0.11 & 0.17 &  \\
CSS080514:162606+225044   & J162605.66+225043.5 & UG     &      & 22.56  & -0.40 & 0.45 & 0.16 & 0.89 &  \\
CSS080426:162657-002549   & J162656.79-002549.3 & UG     &      & 22.24  & -0.32 & 0.31 & 0.49 & 0.03 &  \\
SDSS J162718.39+120435.0  & J162718.39+120435.0 & UGSU   & 2.50 & 19.05  & -0.22 & 0.16 & 0.23 & 0.37 & D \\
V592 Her                  & J163056.39+211658.4 & UGWZ   & 1.34 & 21.45  & 0.21 & 0.00 & -0.11 & 0.10 &  \\
SDSS J163722.21-001957.1  & J163722.21-001957.1 & UG     & 1.62 & 16.59* & 0.23 & 0.01 & -0.15 & -0.09 &  \\
V544 Her                  & J163805.39+083758.3 & UG     & 1.66 & 19.03  & 0.64 & 0.72 & 0.44 & 0.19 &  \\
V610 Her                  & J164339.10+223125.2 & UG     &      & 21.13  & 0.13 & 0.30 & 0.26 & 0.19 & A \\
AH Her                    & J164410.01+251501.9 & UGZ    & 6.19 & 15.00  & -0.52 & -0.02 & 0.65 & 0.72 &  \\
V611 Her                  & J164449.42+195940.0 & UG     &      & 20.70  & -0.17 & -0.03 & 0.10 & 0.55 & D \\
SDSS J170213.26+322954.1  & J170213.25+322954.1 & UGSU+E & 2.40 & 17.83  & 0.24 & 0.09 & 0.42 & 0.42 &  \\
SDSS J173008.38+624754.7  & J173008.35+624754.4 & UGSU   & 1.84 & 16.14* & 0.15 & -0.24 & -0.12 & 0.00 &  \\
SDSS J204448.92-045928.8  & J204448.92-045928.8 & UG     & 40.32 & 16.28  & 0.48 & 0.58 & 0.28 & 0.28 &  \\
SDSS J205914.87-061220.5  & J205914.87-061220.4 & UG     & 1.79 & 18.40  & -0.18 & -0.02 & 0.20 & 0.20 &  \\
SDSS J210014.12+004446.0  & J210014.11+004445.9 & UGSU   & 2.00 & 18.65  & -0.21 & 0.09 & 0.08 & 0.03 & A \\
SDSS J210449.95+010545.9  & J210449.94+010545.8 & UG     & 1.73 & 17.18* & 0.32 & -0.15 & -0.12 & -0.08 &  \\
SDSS J211605.43+113407.5  & J211605.43+113407.2 & UG     & 1.34 & 15.54* & 0.12 & -0.28 & -0.20 & -0.13 & C \\
SDSS J213122.40-003937.0  & J213122.40-003937.0 & UGSU   & 1.52 & 21.44  & 0.24 & -0.06 & -0.22 & 0.42 &  \\
CSS080404:213309+155004   & J213309.41+155004.1 & UG     &      & 21.92  & 0.26 & 0.16 & 0.47 & 0.17 & A \\
CSS071116:214843-000723   & J214842.52-000723.4 & UG     &      & 22.40  & 0.67 & 0.55 & -0.14 & -0.10 &  \\
ROTSE3 J221519.8-003257.2 & J221519.81-003257.2 & UG     &      & 21.38  & -0.25 & 0.31 & 0.13 & 0.51 & A \\
SDSS J223252.35+140353.0  & J223252.34+140352.9 & UG     &      & 17.80* & 0.08 & -0.14 & -0.08 & -0.09 &  \\
SDSS J223439.93+004127.2  & J223439.92+004127.3 & UG     & 2.12 & 17.49* & -0.18 & 0.19 & 0.11 & 0.15 &  \\
SDSS J225831.18-094931.7  & J225831.18-094931.6 & UGSU   & 1.98 & 15.43  & -0.39 & 0.18 & -0.03 & 0.20 &  \\
SDSS J230351.64+010651.0  & J230351.63+010651.1 & UG     & 1.84 & 17.87  & -0.29 & 0.28 & 0.13 & 0.33 &  \\
\hline
\end{tabular}
\end{center}
\end{table*}

\setcounter{table}{2}
\begin{table*}
\begin{center}
\caption{Observed outbursts of the new dwarf novae.  }
\label{OutburstsFull}
\begin{tabular}{cl} 
\hline
SDSS                & Outbursts \\
\hline
J013645.81-193949.1 & Oct 1997 (NOFS), Dec 2000 (NEAT), Nov 2004 (CMC14) \\
J015237.83-172019.3 & Sep 1978 (NOFS) \\
J032015.29+441059.3 & Nov 1994 (NOFS), Dec 2005 (SDSS) \\
J064911.48+102322.1 & Nov 2006 (SDSS) \\
J073208.11+413008.7 & Mar 1953 (NOFS), Dec 2001 (NEAT), Nov 2003 (SDSS), Feb 2007 (GALEX + CRTS) + 3 outburst (CRTS) \\
J073758.55+205544.5 & Jan 2002 (SDSS), Dec 2002 (NEAT), Feb 2006 (CRTS) \\
J074500.58+332859.6 & See text \\
J074859.55+312512.6 & Dec 2001 (SDSS + NEAT) + 10 outbursts (CRTS) \\
J075107.50+300628.4 & Mar 2003 (CMC14 + NEAT), Mar 2005 (CRTS), Jan 2006 (CRTS), Jan 2007 (CRTS), Dec 2007 (CRTS) \\
J075117.00+100016.2 & Nov 1997 (NOFS), Mar 2001 (PM2000), Apr 2005 (CRTS) \\
J075713.81+222253.0 & Jan 1989 (NOFS), Nov 1995 (NOFS), Dec 2003 (GALEX), May 2005 (CRTS), Feb 2006 (CRTS), Dec 2007 (CRTS) \\
J080033.86+192416.5 & Mar 2004 (SDSS) + 9 outbursts in CRTS data \\
J080306.99+284855.8 & Mar 1955 (NOFS), Jan 2002 (SDSS), Nov 2002 (CMC14) + 7 outbursts in CRTS data \\
J081030.45+091111.7 & Dec 2005 (SDSS) \\
J081408.42+090759.1 & Mar 1951 (NOFS), Jan 2001 (NEAT), Apr 2008 (CRTS) \\
J081529.89+171152.5 & Nov 1998 (NOFS) \\
J082648.28-000037.7 & Jan 1991 (NOFS) \\
J083132.41+031420.7 & Feb 2001 (SDSS + NEAT) + 7 outbursts in CRTS data \\
J083508.99+600643.9 & Nov 2003 (SDSS) \\
J084011.95+244709.8 & Apr 2003 (SDSS) \\
J084108.10+102536.2 & Nov 1995 (NOFS), Nov 2007 (CRTS) \\
J091147.02+315101.8 & Jan 2003 (SDSS), Apr 2005 (CRTS), Dec 2005 (CRTS) \\
J091242.18+620940.1 & Jan 1998 (NOFS) \\
J091741.29+073647.4 & Feb 1991 (NOFS) \\
J092620.42+034542.3 & Feb 2004 (GALEX) + 7 outbursts in CRTS data \\
J092809.84+071130.5 & Jan 1996 (NOFS), Nov 2002 (SDSS + NEAT) \\
J093946.03+065209.4 & Oct 1984 (DSS), Mar 2002 (SDSS + NEAT), May 2005 (CRTS), Jun 2008 (CRTS) \\
J100243.11-024635.9 & Apr 1987 (NOFS), Jan 2002 (NEAT), Dec 2005 (UKIDSS + CRTS), Dec 2008 (CRTS) \\
J100516.61+694136.5 & Mar 1953 (NOFS) \\
J105333.76+285033.6 & Feb 2005 (GALEX), highly variable (mag. 17.2-20.6) in CRTS data \\
J112003.40+663632.4 & Apr 2000 (SDSS), Jan 2006 (CRTS) \\
J120054.13+285925.2 & Apr 2002 (NEAT), Dec 2004 (SDSS) \\
J124328.27-055431.0 & Apr 2004 (GALEX) \\
J124602.02-202302.4 & Feb 1978 (NOFS) \\
J124719.03+013842.6 & May 1990 (NOFS), Apr 2004 (GALEX), Jun 2006 (CRTS) \\
J131432.10+444138.7 & May 1990 (NOFS), Mar 2003 (SDSS) \\
J132040.96-030016.7 & May 1983 (NOFS), Apr 2004 (GALEX) \\
J132715.28+425932.8 & Apr 2004 (GALEX) \\
J133820.56+041807.4 & Apr 2001 (SDSS) \\
J141029.09+330706.2 & Apr 2004 (SDSS) \\
J142414.20+105759.8 & Mar 2003 (SDSS) \\
J142953.56+073231.2 & Apr 1989 (NOFS), Apr 2000 (NEAT), Jun 2006 (UKIDSS) + 4 outbursts (CRTS) \\
J151109.79+574100.3 & Jun 2004 (GALEX), Feb 2009 (CRTS) \\
J152124.38+112551.9 & Jun 2003 (SDSS) \\
J153457.24+505616.8 & Jun 1954 (NOFS), Apr 2007 (GALEX) \\
J154357.66+203942.1 & Mar 1992 (NOFS), May 2003 (NEAT), May 2004 (SDSS) + 7 outbursts in CRTS data \\
J154652.70+375415.2 & May 2003 (SDSS), Mar 2006 (CRTS), Sep 2006 (CRTS), Jun 2007 (CRTS), Mar 2009 (CRTS) \\
J154817.56+153221.2 & Feb 1999 (2MASS), Apr 2005 (CRTS), Jul 2006 (GALEX + CRTS) \\
J155030.38-001417.3 & May 1987 (NOFS), Aug 1988 (NOFS), Mar 1994 (NOFS), Jul 2007 (GALEX) + 7 outbursts in CRTS data \\
J155540.19+364643.1 & Jun 1988 (NOFS), Mar 2002 (CMC14), Jun 2005 (CRTS), May 2007 (CRTS), May 2008 (CRTS), Feb 2009 (CRTS) \\
J161027.61+090738.4 & Jun 1998 (NEAT) \\
J161442.43+080407.9 & Jul 2004 (CMC14) \\
J162520.29+120308.7 & Aug 2001 (NEAT) \\
J162558.18+364200.6 & Apr 1994 (NOFS) \\
J162900.55+341022.0 & Jun 1954 (NOFS) \\
J164705.07+193335.0 & May 1992 (NOFS) \\
J170145.85+332339.5 & Apr 2001 (SDSS) \\
J170810.31+445450.7 & Jul 1990 (NOFS) \\
J171202.95+275411.0 & Jun 1950 (NOFS), May 2001 (SDSS) + 6 outbursts in CRTS data \\
\hline
\end{tabular}
\end{center}
\end{table*}

\begin{table*}
\begin{center}
\contcaption{}
\begin{tabular}{cl} 
\hline
SDSS                & Outbursts \\
\hline
J174839.77+502420.3 & Oct 2002 (SDSS) \\
J191616.53+385810.6 & May 1955 (NOFS), Jul 1988 (NOFS) \\
J205931.86-070516.6 & Aug 1953 (NOFS), Oct 2008 (CRTS) \\
J212025.17+194156.3 & Jul 1954 (NOFS), Jul 2002 (NEAT), Sep 2006 (GALEX + CRTS) \\
J223854.51+053606.8 & Nov 1983 (DSS) \\
\hline
\end{tabular}
\end{center}
\end{table*}

\label{lastpage}

\end{document}